\documentclass[pra, aps,showpacs,twocolumn,floatfix,superscriptaddress,amsmath]{revtex4-1}
\usepackage{graphicx}
\usepackage{graphics}
\usepackage{mathtools}
\usepackage{float}
\usepackage{amssymb}
\def \bk{{\bf k}}
\def \br{{\bf r}}
\def \bv{{\bf v}}
\def \be{{\bf e}}

\def \cH{{\cal{H}}}

\begin{document}

\title{Probing superfluidity of Bose-Einstein condensates via laser stirring}
\author{Vijay Pal Singh}
\affiliation{Zentrum f\"ur Optische Quantentechnologien, Universit\"at Hamburg, 22761 Hamburg, Germany}
\affiliation{Institut f\"ur Laserphysik, Universit\"at Hamburg, 22761 Hamburg, Germany}
\affiliation{The Hamburg Centre for Ultrafast Imaging, Luruper Chaussee 149, Hamburg 22761, Germany}
\author{Wolf Weimer}
\affiliation{Institut f\"ur Laserphysik, Universit\"at Hamburg, 22761 Hamburg, Germany}
\author{Kai Morgener}
\affiliation{Institut f\"ur Laserphysik, Universit\"at Hamburg, 22761 Hamburg, Germany}
\affiliation{The Hamburg Centre for Ultrafast Imaging, Luruper Chaussee 149, Hamburg 22761, Germany}
\author{Jonas Siegl}
\author{Klaus Hueck}
\affiliation{Institut f\"ur Laserphysik, Universit\"at Hamburg, 22761 Hamburg, Germany}
\author{Niclas Luick}
\author{Henning Moritz}
\affiliation{Institut f\"ur Laserphysik, Universit\"at Hamburg, 22761 Hamburg, Germany}
\affiliation{The Hamburg Centre for Ultrafast Imaging, Luruper Chaussee 149, Hamburg 22761, Germany}
\author{Ludwig Mathey}
\affiliation{Zentrum f\"ur Optische Quantentechnologien, Universit\"at Hamburg, 22761 Hamburg, Germany}
\affiliation{Institut f\"ur Laserphysik, Universit\"at Hamburg, 22761 Hamburg, Germany}
\affiliation{The Hamburg Centre for Ultrafast Imaging, Luruper Chaussee 149, Hamburg 22761, Germany}
\date{\today}

\begin{abstract}
We investigate the superfluid behavior of a Bose-Einstein condensate of $^{6}$Li molecules. In the experiment by Weimer {\it et al.}, Phys. Rev. Lett. \textbf{114}, 095301 (2015) a condensate is stirred by a weak, red-detuned laser beam along a circular path around the trap center. The rate of induced heating increases steeply above a velocity $v_c$, which we define as the critical velocity. Below this velocity, the moving beam creates almost no heating. In this paper, we demonstrate a quantitative understanding of the critical velocity. Using both numerical and analytical methods, we identify the non-zero temperature, the circular motion of the stirrer, and the density profile of the cloud as key factors influencing the magnitude of $v_c$. A direct comparison to the experimental data shows excellent agreement. 
\end{abstract}
\maketitle

\section{Introduction}
 The intriguing phenomenon of superfluidity arises from an interplay of quantum statistics and interactions at low temperatures. Its defining feature is the stability against external perturbations and, closely related, dissipationless flow near an obstacle. However, the superfluid behavior will only be sustained within a certain parameter regime. If the system is perturbed by a sufficiently strong or sufficiently fast moving 
 perturbation, its dissipationless nature will break down.
   The onset of dissipation typically occurs around a velocity which is referred to as the critical velocity.


In a seminal study, Landau estimated the critical velocity by considering the onset of dissipation due to the excitation of single  elementary quasi-particles \cite{Landau1941} of the system. 
In order to satisfy both energy and momentum conservation, 
an elementary excitation of energy $\epsilon(\bk)$ with momentum $\hbar \bk$ can only be created above the velocity $v_c= \mathrm{min}_\bk   [ \epsilon(\bk)/(\hbar k) ]$, which is the Landau criterion. 
$\hbar$ is the Planck constant and  $\bk$ is the wave vector, with $k = |\bk|$. 
When this expression is applied to the Bogoliubov approximation of the spectrum of a weakly interacting Bose gas, it predicts the sound velocity $v_{\mathrm{B}}$ as the critical velocity. This is the natural scale for the critical velocity, if the decay mechanism is due to phononic excitations.
In addition to phonons, however, other types of excitations can control the critical velocity. In Ref. \cite{Feynman1955}, 
Feynman considered the superfluid flow into a long channel with a diameter $D\gg \xi$ \cite{Fetter1976}, and estimated a critical velocity of $v_c \approx \hbar/(mD)\ln(D/\xi)$. $\xi$ is the healing length and $m$ is the atomic mass. Thus, vortex excitations can result in a lower critical velocity than phonon excitations, if this velocity scale is smaller than the phonon velocity.

\begin{figure}
\includegraphics[width=1.0\linewidth]{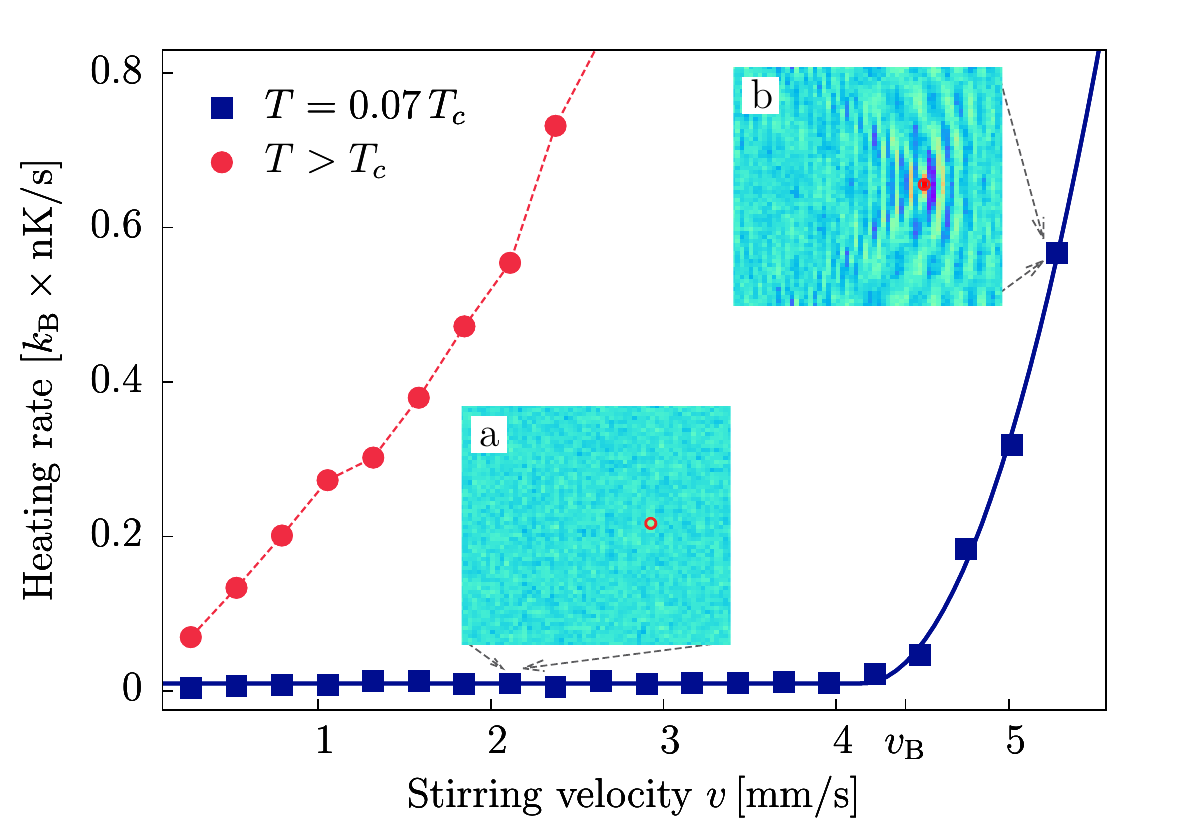}
\caption{(Color online) Simulated heating rates for a homogeneous condensate at $T=0.07\, T_c$ (blue squares), stirred along a linear path by a weak stirrer of strength $V_0= -0.03\, \mu$. A critical velocity of $4.2\, \mathrm{mm/s}$ is determined using the fitting function given by Eq. \ref{eq_fit_function}. The fitted curve is shown by the solid line. 
The Bogoliubov estimate of the sound velocity $v_{\mathrm{B}}$ is $4.4\, \mathrm{mm/s}$.
The insets are snap shots of the density: In panel (a) the moving perturbation causes no drag for $v<v_c$, whereas panel (b) demonstrates the formation of sound waves leading to heating for $v>v_c$. The stirrer location is marked by a circle of radius $\sigma$. 
For the thermal gas  with $T>T_c$ (red circles) heating increases linearly in $v$. Here, we use $V_0=-0.35\, \mu$. }
\label{fig_idealized}
\end{figure}

  Several experiments have probed the superfluidity of dilute Bose gases via a local perturbation, in particular via laser stirring. In Ref. \cite{Ketterle1999} a three dimensional (3D) condensate and in Ref. \cite{Dalibard2012} a two dimensional condensate were stirred with a blue-detuned laser beam, moving on a linear and circular trajectory, respectively. The breakdown of superfluid flow due to a constriction was probed in 
Ref. \cite{Ramanathan2011}: a toroidal Bose-Einstein condensate (BEC) was put into quantized rotation, and the decay due to a potential barrier, created via  a blue-detuned laser, was observed. It was demonstrated in Ref. \cite{Mathey2014}, that the onset of heating is governed by thermally activated phase slips, due to vortices traversing the barrier region. Recently, superfluidity was observed in an elongated $^6$Li gas oscillating with respect to a $^7$Li BEC \cite{Salomon2014}. Here, the onset of heating is predicted to occur for a relative velocity that equals the sum of the individual sound velocities \cite{Salomon2015}.     
In addition to these studies, superfluidity in BECs has been studied in a number of theoretical works based on the Gross-Pitaevskii equation  \cite{Winiecki1999, Stiessberger2000, Jackson2000, Crescimanno2000}.   Further studies were reported in Refs. \cite{theoryrefs}.

As a new, and previously unexplored feature of stirring experiments, Ref. \cite{Weimer2015} reported on perturbing a cloud of $^{6}$Li atoms and molecules, across the BEC-BCS cross-over, with a red-detuned laser, resulting in an attractive stirring potential.
In this paper, we demonstrate a quantitative understanding of the superfluid response of the condensate of $^{6}$Li molecules.
We first demonstrate that the seemingly minor detail of stirring with a red-detuned laser, rather than a blue-detuned one, results in a qualitatively different scenario for the breakdown of superfluidity. We show that for a blue-detuned laser, and for intermediate laser intensity, the primary dissipation mechanism consists of shedding vortex--anti-vortex pairs, rather than fully deconfined vortices, as in the Feynman scenario.
   Interestingly, for a red-detuned laser, neither vortices nor vortex--anti-vortex pairs are generated. Therefore, the choice of a red-detuned laser, with weak or intermediate intensity, compared to the mean-field energy, allows to study the phononic dissipation mechanism with clarity.
We identify the condensate temperature, the circular motion of the stirrer which was used in Ref.  \cite{Weimer2015}, and the inhomogeneous density of the cloud as key parameters that influence the magnitude of $v_c$. These factors reduce the critical velocity below the phonon velocity, even though the dissipation mechanism is phononic.
  A comparison of our  numerical and analytical results, and the experimental data of Ref. \cite{Weimer2015} in the BEC regime, gives good agreement, and develops a consistent physical picture.

This paper is organized as follows: In Sec. \ref{sec_sim_method} we describe the simulation method that we use. In Sec. \ref{sec_heating_rate} we develop an analytical expression for the heating rate within the Bogoliubov approximation, for linear stirring, and compare this result to the simulation in the corresponding regime. 
In Sec. \ref{sec_att_vs_rep} we study the critical velocity for a repulsive stirring potential and contrast the resulting dissipation mechanism to that of an attractive stirring potential. 
In Sec. \ref{sec_circ_vs_lin} we develop an analytical expression  for stirring along a circular path, and compare this result to the simulation.
 In Sec. \ref{sec_vc_temp} we discuss the reduction of $v_c$ for finite temperatures.   
In Sec. \ref{sec_trapped_system} we identify the influence of the inhomogeneous density distribution of the molecular cloud on $v_c$.
In Sec. \ref{sec_comparison} we compare the numerical and analytical results to the experiment, and  in Sec. \ref{sec_conclusion} we conclude.

\begin{figure*}[]
\includegraphics[width=0.95\linewidth]{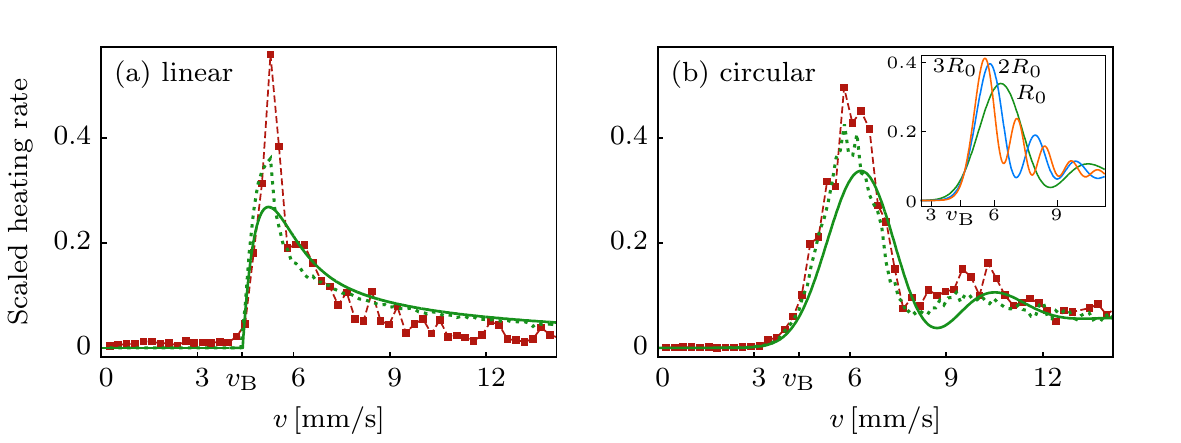}
\caption{(Color online) 
  We show the numerically determined heating rate (red squares connected with a dashed line), that is induced by stirring with a weak stirring potential of strength $V_0\approx -0.03\, \mu$, in comparison to the analytical prediction (green, continuous lines).
In panel (a), we show the case of a linear stirring motion. 
In panel (b) we show the heating rate induced by a circular stirring motion as a function of the stirring velocity, for a stirring  radius of $R_0=10\, \mu \mathrm{m}$. 
In addition to the analytical prediction for the continuum case, we also show the heating for a lattice system (green, dotted lines). The inset of panel (b) shows the predictions for three different radii, in particular $10\, \mu \mathrm{m}$, $20\, \mu \mathrm{m}$, and $30\, \mu \mathrm{m}$. As the radius is increased, the heating rate due to a circular motion slowly recovers the heating rate for a linear motion.
 Overall,  the analytical results capture the numerical results well.  }
\label{fig_circ_lin_comp}
\end{figure*}

\section{Simulation method} \label{sec_sim_method}
We simulate the stirring dynamics of the system within a c-field formalism. 
The Hamiltonian of the unperturbed system is
\begin{align} \label{eq_hamil}
\hat{H}_{0} &= \int \mathrm{d}{\bf r} \Big[ - \frac{\hbar^2}{2m}  \nabla \hat{\psi}^\dagger({\bf r}) \cdot \nabla \hat{\psi}({\bf r})  + V({\bf r}) \hat{\psi}^\dagger({\bf r})\hat{\psi}({\bf r}) \nonumber\\
&   + \frac{g}{2} \hat{\psi}^\dagger({\bf r})\hat{\psi}^\dagger({\bf r})\hat{\psi}({\bf r})\hat{\psi}({\bf r})\Big],
\end{align}
where $\hat{\psi}$ $(\hat{\psi}^\dagger)$ is the field annihilation (creation) operator, and $g=4\pi a_s \hbar^2/m$ is the interaction constant; $a_s$ is the $s$-wave scattering length. The external potential $V({\bf r})$ describes the harmonic trap, $V_{\mathrm{trap}}({\bf r})=m(\omega_r^2r^2+\omega_z^2z^2)/2$. $\omega_r$ and $\omega_z$ are the trap frequencies in the radial and transverse directions, respectively, and $r=(x^2+y^2)^{1/2}$ is the radial coordinate. 
We add the following term to describe laser stirring
\begin{align} \label{eqn_hamiltonian_stir}
\hat{ \mathcal{H}}_{s}(t) = \int \mathrm{d}{\bf r}\, V({\bf r},t) \hat{n}({\bf r}),
\end{align} 
where $V({\bf r},t)$ is the time-dependent stirring potential, $\hat{n}({\bf r})$ is the density operator at the location  ${\bf r}=(x,y,z)$.
The stirring potential is a Gaussian along the $x$ and $y$ direction with a width $\sigma$, and independent of the $z$ direction:
\begin{equation}\label{stirringpot} 
V({\bf r},t)  = V_0 \exp \Bigl(- \frac{ \bigl(x -x_s(t) \bigr)^2 + \bigl(y -y_s(t) \bigr)^2}{2\sigma^2} \Bigr),
\end{equation}
with a strength $V_0$. It is centered around $x_s(t)$ and $y_s(t)$, which move either along a linear or a circular path as a function of time $t$.

To simulate this system numerically, we map it on a lattice system, which also introduces a short range cut-off, see Appendix \ref{App_sim_eRate}. 
 We represent both the equations of motion and the initial state within a c-number representation, which corresponds to formally replacing the operators $\hat{\psi}$ by complex numbers. Furthermore, we approximate the initial ensemble by a classical ensemble, rather than a non-classical phase-space distribution, within a grand canonical ensemble of temperature $T$ and chemical potential $\mu$. This can also be seen as the classical limit of the truncated Wigner approximation, as discussed and applied in Ref. \cite{Mathey2014}, which also corresponds to molecular dynamics in the classical wave limit.
The initial states are generated via a classical Metropolis algorithm.

Motivated by the experimental setup of Ref. \cite{Weimer2015}, 
we consider a pancake-shaped, three-dimensional condensate. The density distribution resembles an oblate spheroid. 
We employ a lattice with $140\times140\times11$ sites. 
For comparison, we consider a system with homogeneous density, without a trapping potential $V(\br)$, for which we use a lattice with $60\times60\times3$ sites. 
For the discretization length $l$ we choose $1\, \mu\mathrm{m}$. As discussed in Ref. \cite{Mora2003}, $l$ is chosen such that the healing length $\xi \equiv\hbar/\sqrt{2mgn}$ and the thermal de Broglie wavelength, $\lambda=\sqrt{2\pi\hbar^2/mk_{\mathrm{B}}T}$ are comparable to, or larger than $l$, $n$ being the density, and $k_{\mathrm{B}}$ the Boltzmann constant. In Ref. \cite{Weimer2015} the condensate is stirred using a narrowly focused red-detuned laser beam that forms an attractive Gaussian potential, whose width is of the order of the healing length $\xi$. 
  For simulations of this system we adjust the parameters according to the experimental choices, in particular we adjust the central column density, the scattering length $a_s$, the temperature, the stirring time, the stirrer strength $V_0$, and the stirring pattern. Typical trap frequencies are $\omega_r \approx 2\pi \times 30 \, \mathrm{Hz}$ and $\omega_z \approx 2\pi \times 550 \, \mathrm{Hz}$.
  For simulations of the homogeneous system we use $\sigma = 1.1\, \mu\mathrm{m}$,  $n=0.48\, \mu\mathrm{m}^{-3}$, and $a_s=2180a_0$, where $a_0$ is the Bohr radius. These parameters are in the typical range of the experimental parameters. 
  The stirring sequence is the following: We linearly turn on the stirring potential over $10 \, \mathrm{ms}$, let it stir the system for $200 \, \mathrm{ms}$, and then turn it off over $5 \, \mathrm{ms}$. This is again inspired by the experimental choices. 
    We repeat this for various stirring velocities $v$ and calculate a thermal ensemble of the total energy $E = \langle \hat{H}_{0}\rangle$ using the unperturbed Hamiltonian in Eq. \ref{eq_hamil}. During the stirring process the energy increases linearly. From this linear increase, we determine the heating rate ${dE/dt}$ for each $v$, and for the desired parameter regimes. We elaborate on the numerical method of determining the heating rate in Appendix \ref{App_sim_eRate}.

\section{Homogeneous BEC with linear stirring} \label{sec_heating_rate}
While the experimental setup displays an interplay of features which complicate the interpretation, such as the temperature, the stirring pattern and the density inhomogeneity, we isolate these features in our analysis to identify how they affect the heating rate.
   We start out with the case of  a homogeneous condensate at a low temperature, stirred along a linear path with a weak stirrer. 
      In Fig. \ref{fig_idealized} we show the simulated heating rate as a function of the stirring velocity $v$, for a condensed and a thermal system. For the condensed system, the moving stirrer creates almost no dissipation at low velocities $v$. 
    For velocities above the sound velocity $v_{\mathrm{B}}$, where $v_{\mathrm{B}} \equiv \sqrt{gn/m}$ is the Bogoliubov estimate, sound-wave excitations are created and the rate of induced heating increases steeply. For these conditions, in particular for a temperature much smaller than the mean-field energy and for a weak, linearly moving stirrer, we recover the estimate $v_{c}\approx  v_{\mathrm{B}}$.
   For comparison, we  show the heating rate of a stirred, thermal gas at a temperature $T>T_c$ in Fig. \ref{fig_idealized}. For this regime, we recover a linear dependence, corresponding to friction that scales linear in $v$. 
   
   Throughout this paper, to quantify the critical velocity, we use the following fitting function of the heating rate
\begin{equation} \label{eq_fit_function}
\Bigl( \frac{dE}{dt}\Bigr)_{\mathrm{Fit}} = A\frac{(v^2-v_c^2)^2}{v} + B
\end{equation}  
 which was discussed  in Ref. \cite{Pitaevskii2004}, for $v>v_c$, with the free parameters $A$, $B$, and $v_c$. This fitting procedure gives a robust estimate of $v_{c}$ for both numerical and analytical results, even though the analytical form that we find differs from this expression.
 For the heating rate of the condensed system in Fig. \ref{fig_idealized}, we determine a critical velocity of $4.2\, \mathrm{mm/s}$, which is comparable to $v_{\mathrm{B}} =4.4\, \mathrm{mm/s}$.

  In addition to the numerical results in this paper, we develop an analytical expression for the heating rate that quantitatively describes the simulation results. We derive the heating rate using the stirring potential as a perturbation Hamiltonian that we linearize within the Bogoliubov approximation. We extend this to finite temperatures in Sec. \ref{sec_vc_temp}, where we include the thermal broadening of the phonon modes. 
  
  In a general setting, not necessarily for a condensate or even bosons, we determine the heating rate for a weak stirrer perturbatively in the stirring term:
\begin{align}
\frac{d \langle \hat{H}_{0}(t)\rangle}{ d t} &= \frac{i}{\hbar} \langle [\hat{H}_{s,I}(t), \hat{H}_{0}] \rangle\nonumber\\
&  - \frac{1}{\hbar^2}\int_{0}^{t} d t_{1} \langle [\hat{H}_{s,I}(t_{1}), [\hat{H}_{s,I}(t), \hat{H}_{0}]]\rangle \ldots\label{rate}
\end{align}
where we gave the perturbative expansion to second order. 
$\hat{H}_{0}$ is the Hamiltonian of the unperturbed system, such as Eq. \ref{eq_hamil}, and therefore a conserved quantity.  $\hat{H}_{s,I}(t)$ is the perturbation in the interaction picture
\begin{equation}
\hat{H}_{s,I}(t) = \exp(i \hat{H}_{0} t) \hat{H}_{s}(t) \exp(- i \hat{H}_{0} t).
\end{equation}
$\hat{H}_{s}(t)$ is the stirring term in the Schr\"odinger picture, such as Eq. \ref{eqn_hamiltonian_stir}.
The expectation value $\langle \ldots \rangle$ is taken with regards to a density operator $\rho$, i.e. $\langle \ldots \rangle = \mathrm{Tr} (\rho \ldots)$. 
If $\rho$ commutes with $\hat{H}_{0}$, as it does for a canonical ensemble, the first order term in Eq. \ref{rate} vanishes.

\begin{figure}[]
\includegraphics[width=1.0\linewidth]{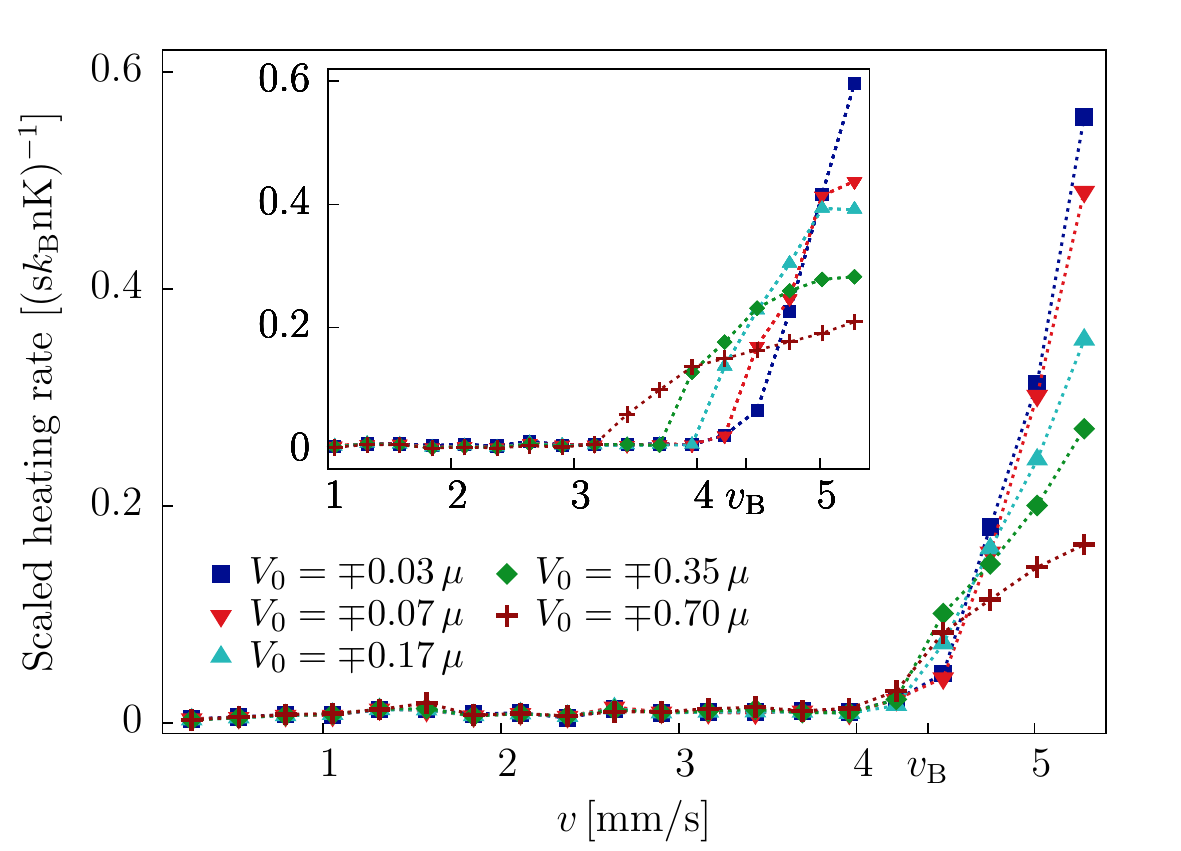}
\caption{(Color online) Simulated heating rates for attractive and repulsive stirrers are obtained for various stirrer strengths $V_0$. The negative $V_0$ stands for attractive stirrer, whereas the positive $V_0$ for repulsive stirrer. The inset shows the heating rates for the repulsive stirrer.}
\label{fig_att_rep}
\end{figure}

We now evaluate this expression for a linear stirring motion, also see \cite{remark}.
We  consider a stirring potential as in Eq. \ref{stirringpot}, with a linear stirring trajectory $\br_{s}(t) = \bv t$, with $\br_{s}(t)=(x_{s}(t), y_{s}(t))$ and ${\bf v} = (v_x, v_y)$. 
In momentum representation, the stirring term is
\begin{equation}
\hat{\cH}_{s}(t) = \sum_{\bk} V_{\bk}(t) \hat{n}_{\bk}.
\end{equation}
$V_{\bk}$ is the Fourier transform of the stirring potential with $\bv = 0$, in particular $V_\bk =  2\pi V_0 \sigma^2/A \times \delta_{k_z} \exp(-k^2\sigma^2/2)$ with $A$ being the system area. $V_{\bk}(t)$ is the moving stirring potential, which is $V_{\bk}(t) = V_{\bk} \exp(i \bk \bv t)$.
$\hat{n}_{\bk}$ is the Fourier transform of the density operator. We use the Bogoliubov approximation to evaluate the heating rate. So the Hamiltonian is approximately
\begin{equation}\label{HamBogoliubov}
\hat{\cH}_{0} = \sum_{\bk} \hbar \omega_{k} \hat{b}_{\bk}^{\dagger} \hat{b}_{\bk},
\end{equation}
where $\hat{b}_{\bk}$, $\hat{b}^\dagger_{\bk}$ are the Bogoliubov operators, and $\hbar \omega_{k}$ is the Bogoliubov dispersion, $\hbar \omega_{k} = \sqrt{\epsilon_k (\epsilon_k + 2 g n_c)}$, where $\epsilon_k$ is the dispersion of the non-interacting system, and $n_c$ is the condensate density.
The density operator is $\hat{n}_{\bf k} \approx \sqrt{N_0}(u_k + v_k) (\hat{b}_{-{\bf k}}^\dagger + \hat{b}_{\bf k})$ with $N_0$ being the number of condensed atoms. $u_k$, $v_k$ are the Bogoliubov functions, with $(u_{k} + v_{k})^{2} = \epsilon_k/(\hbar \omega_{k})$. 
With this, $\hat{\cH}_{s}(t)$ is
\begin{equation} \label{stirBogoliubov} 
\cH_{s}(t) \approx \sum_{\bk} V_{\bk}(t)  \sqrt{N_{0}} (u_{k} + v_{k}) (\hat{b}_{-{\bf k}}^\dagger + \hat{b}_{\bf k}). 
\end{equation}
$\hat{\cH}_{s, I}(t)$ has the same form with $\hat{b}_{\bk}  \rightarrow \hat{b}_{\bk}\exp(- i \omega_{k} t)$ and $\hat{b}_{-\bk}^{\dagger}  \rightarrow  \hat{b}_{-\bk}^{\dagger} \exp(i \omega_{k} t)$. 
With this, we obtain
\begin{align}
&  - \int_{0}^{t} d t_{1} \langle [\cH_{s}(t_{1}), [\cH_{s}(t), \cH_{0}]]\rangle\\
& = 2 \sum_{\bk} |V_{\bk}|^{2}
\hbar \omega_{k} N_{0} (u_{k} + v_{k})^{2}  \frac{\sin((\omega_{k} - \bk \bv) t )}{\omega_{k} - \bk \bv}
\end{align}
which at long times approaches
\begin{equation} \label{eg_heatingRate}
\frac{d E}{dt} = \frac{2\pi}{\hbar} \sum_{\bf k} \omega_k (u_k + v_k)^2 N_0 |V_\bk|^2  \delta(\omega_k - {\bf vk}).
\end{equation}
We note that this result can also be obtained by solving the dynamical evolution that is created by Eqs. \ref{HamBogoliubov} and \ref{stirBogoliubov} exactly, which is 
\begin{equation} \label{eq_ansatz}
\hat{b}_{\bf k}(t) = e^{-i\omega_k t} \hat{b}_{\bf k} + A_{\bf k}(t),
\end{equation}
with
\begin{align} \label{eq_Ak(t)}
A_{\bf k}(t) &= \frac{-2i}{\hbar} (u_k + v_k) \sqrt{N_0}V_\bk \frac{\sin\bigl[ (\omega_k - {\bf vk})t/2 \bigr] }{(\omega_k - {\bf vk})} \nonumber \\
&  \quad \times e^{-i(\omega_k - {\bf vk})t/2}.
\end{align} 
Therefore the energy at time $t$ compared to the initial energy is
\begin{align} \label{eq_energySecond}
\langle \hat{\cH}_{0}(t)\rangle - \langle \hat{\cH}_{0}(0)\rangle &= \frac{4}{\hbar^2} \sum_{\bf k} \hbar\omega_k (u_k + v_k)^2 N_0 |V_\bk|^2  \nonumber \\ 
&  \quad \times \frac{\sin^2\bigl[ (\omega_k - {\bf vk})t/2 \bigr] }{(\omega_k - {\bf vk})^2}. 
\end{align}
From this, we recover  the heating rate given in Eq. \ref{eg_heatingRate}. 
 It is a sum of Fermi's Golden rule terms multiplied by the Bogoliubov excitation energy.
 We use $(u_k + v_k)^2 = \hbar k^2/(2m\omega_k)$ and the explicit form of $|V_\bk|^2$, which results in
\begin{equation} \label{eg_heatingRate2}
 R_\mathrm{sc} = \frac{\pi \sigma^4}{m A} \int \mathrm{d}^2k \, k^2 \exp(-k^2\sigma^2) \delta(\omega_k - vk_x),
\end{equation}
where $k^2=k_x^2+k_y^2$. We defined  the rescaled heating rate $R_\mathrm{sc} \equiv (dE/dt)/(N_0V_0^2)$. 
Without loss of generality, we have set the direction of the stirring velocity along the x-axis,  $\bv = v \be_{x}$.
In addition to this expression for the continuum system, we also evaluate and rescale the expression for the lattice system. 
We use $\epsilon_k = 2J\sum_{j=1}^d (1-\cos(k_j l))$ and numerically evaluate Eq. \ref{eg_heatingRate} for the first Brillouin zone.  $d$ is the spatial dimension, $J=\hbar^2/(2ml^2)$ is the energy, and $l$ is the discretization length. We show the heating rates and their comparison to the simulated rates as a function of $v$ in Fig. \ref{fig_circ_lin_comp}(a). 
\begin{figure}[]
\includegraphics[width=0.95\linewidth]{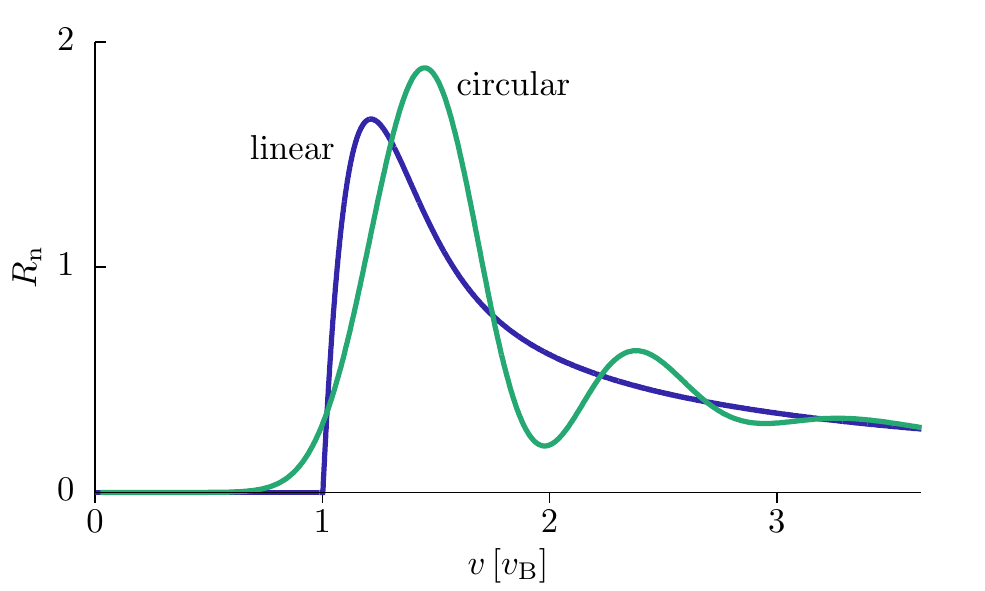}
\caption{(Color online) Analytical heating rate $R_\mathrm{n}$ for linear stirring, Eq. \ref{eq_aRate_norm3} and circular stirring with radius $R_0= 10\, \mu \mathrm{m}$, Eq. \ref{heatingratecirc}, at zero temperature, within the Bogoliubov estimate. The stirring velocity $v$ is shown in units of $v_\mathrm{B}$. }
\label{fig_analytic_circ_lin}
\end{figure}
\begin{figure*}[]
\includegraphics[width=1.00\linewidth]{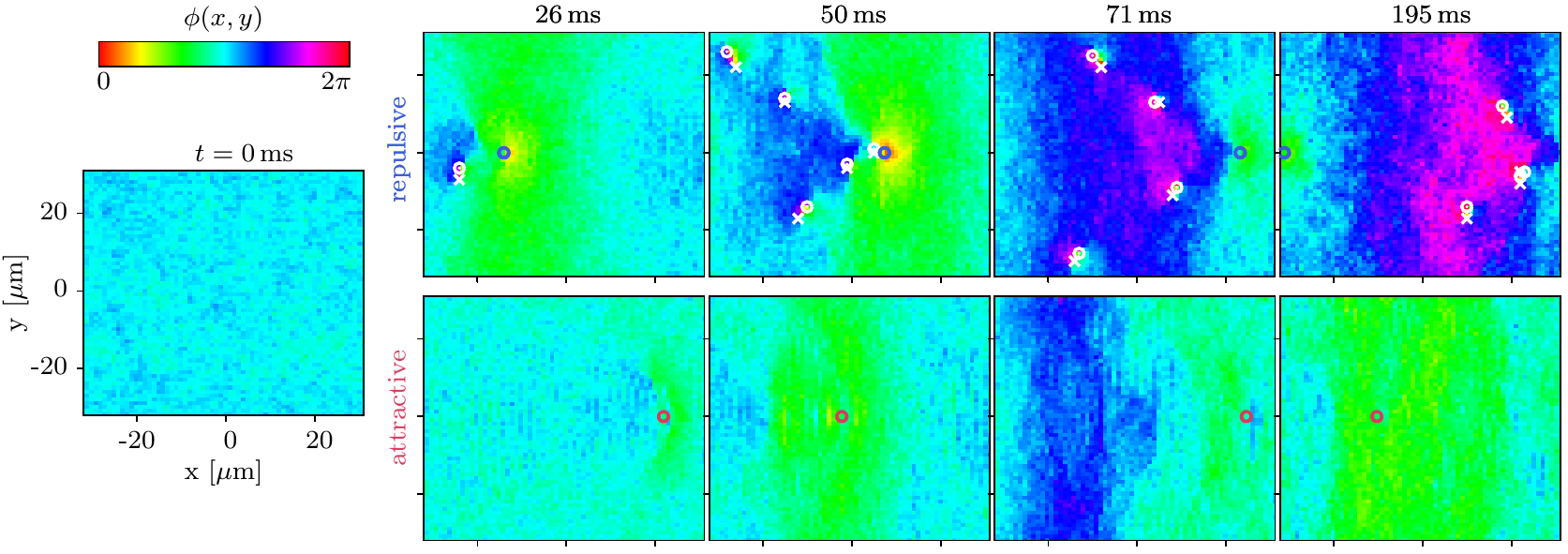}
\caption{(Color online) The phase evolution of a single realization for a repulsive stirrer (upper row) and an attractive stirrer (lower row), at the times $t=26, \, 50,\, 71$, and $195\, \mathrm{ms}$. The location of the stirrer is marked by a circle of radius $\sigma$. For the repulsive stirrer the phase winding around the plaquettes is also indicated. A vortex (circled dot) and an anti-vortex (cross) correspond to a phase winding of  $+2\pi$ and $-2\pi$, respectively. 
 For the repulsive stirrer, dissipation is caused by the creation of  tightly bound vortex pairs, for the attractive stirrer by the creation of phonons. }
\label{fig_vortices_rep}
\end{figure*}

In the continuum limit, the heating rate in Eq. \ref{eg_heatingRate} can be evaluated explicitly as follows: 
We write the integral in polar coordinates, and integrate out the polar angle. This gives
\begin{equation} \label{eq_heatingRateBog}
R_\mathrm{sc} =  \frac{2\pi \sigma^4}{m A}  \int_0^{k_u} \mathrm{d}k\,  \frac{k^3 \exp(-k^2\sigma^2)}{\sqrt{v^2k^2-\omega_k^2}}
\end{equation}
for $v > v_{\mathrm{B}}$, and $R_\mathrm{sc} =0$ else. The wavevector $k_u$ corresponds to the maximal phonon momentum that can be created by the stirring process, it is given by $k_u=\xi^{-1} \sqrt{2(v^2/v_{\mathrm{B}}^2 -1)}$. 
We write the heating rate in dimensionless form, by defining $R_\mathrm{n} \equiv \hbar (dE/dt)/(N_{\mathrm{cyl}}V_0^2)$. 
$N_{\mathrm{cyl}}= \pi N_0 \sigma^2/A$ is the number of atoms in a cylinder of length $L_z$ and radius $\sigma$. 
With  $k=\tilde{k}/\sigma$, the heating rate is
\begin{equation} \label{eq_aRate_norm2}
 R_\mathrm{n} =  \sqrt{8}  \int_0^{\sqrt{2X}} \mathrm{d} \tilde{k}\,  \frac{ \tilde{k}^2 \exp(-\tilde{k}^2)}{\sqrt{X - \tilde{k}^2/2}}
\end{equation}
with $X  \equiv (v^2/v_{\mathrm{B}}^2-1)(\sigma^2/\xi^2)$. The integration can be performed analytically, giving 
\begin{equation} \label{eq_aRate_norm3}
 R_\mathrm{n} =  2 \pi X \exp(-X) \Bigl( I_0(X) - I_1(X) \Bigr),
\end{equation}
where $I_0(x)$ and $I_1(x)$ are the modified Bessel functions of the first kind of order $0$ and $1$, respectively. This function is shown in Fig. \ref{fig_analytic_circ_lin}. 
 The onset of dissipation occurs at $v= v_{\mathrm{B}}$, with a steep, linear slope:
\begin{equation}
R_\mathrm{n} \approx  2 \pi (\sigma^2/\xi^2) (v^2/v_{\mathrm{B}}^2 -1)
\end{equation}
for $v/v_{\mathrm{B}} \gtrsim 1$. For large velocities, $v/v_{\mathrm{B}} \gg 1$, the heating rate falls off as 
\begin{align}
R_\mathrm{n} \approx  \sqrt{\pi/2} (\xi/\sigma) \frac{1}{\sqrt{v^2/v_{\mathrm{B}}^2 -1}}.
\end{align}
The heating rate undergoes a maximum at 
\begin{equation}
v_\mathrm{max}^2 \approx v_{\mathrm{B}}^2( 1+ 0.79 \xi^2/\sigma^2),
\end{equation}
where it assumes the value $(R_\mathrm{n})_\mathrm{max} \approx 1.65$.

\begin{figure}[]
\includegraphics[width=0.60\linewidth]{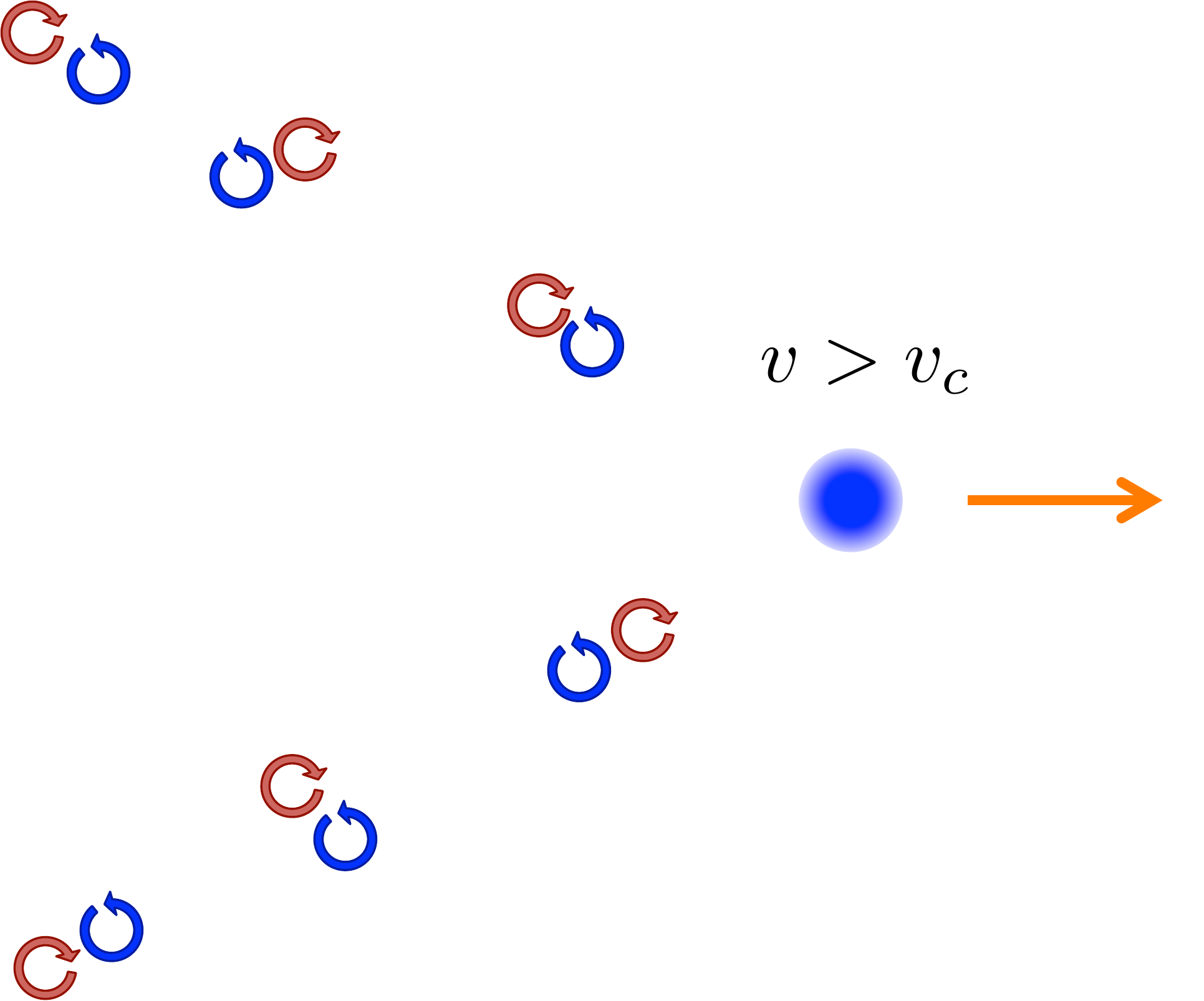}
\caption{(Color online) Illustration of the dissipation mechanism of a repulsive stirrer, due to the creation of  vortex-anti-vortex pairs. The stirrer is depicted by the blue disc, the arrow indicates the direction of the moving stirrer. A vortex (an anti-vortex) is depicted by the anticlockwise (clockwise) open circle arrow. }
\label{fig_sketch_vortexpair}
\end{figure}

\section{Attractive versus repulsive stirrers} \label{sec_att_vs_rep}
In the experiment reported in Ref. \cite{Weimer2015}, the stirring potential was attractive, in contrast to previous measurements such as Refs. \cite{Ketterle1999, Dalibard2012}. 
This superficially minor difference results in a qualitatively different dissipation mechanism as the stirring potential is increased from small magnitudes to the larger ones of the order of the mean-field energy. 
In Fig. \ref{fig_att_rep} we show the heating rate as a function of the stirring velocity for increasing stirring potentials, at a low temperature of
$T=0.07\, T_c$, and for a linear stirring motion. 
For small stirring potentials, compared to the mean-field energy $\mu=gn$, the heating rate for attractive and repulsive potentials agree. For this limit the response is quadratic in the potential, as demonstrated in the previous section, and therefore independent of the sign.
As the magnitude is increased for the attractive potential, the velocity at which the dissipation increases steeply is only slightly reduced. But for the repulsive stirring potential a visible reduction is achieved, indicating the appearance of an additional dissipative mechanism.


To understand the additional decay mechanism for the repulsive stirring potential, we show the time evolution of the phase field $\phi(\br, t)$ of a single realization of the ensemble that is used in the simulation.
This field is derived from the complex field $\psi ({\bf r})$ via the phase-density representation $\psi = \sqrt{n}\exp(i\phi)$. 
In Fig. \ref{fig_vortices_rep}, we show the time evolution of the phase field for a single $xy$-plane of the three dimensional lattice, for both an attractive and a repulsive stirring potential. The magnitude of the stirrer is $V_0=\pm0.70 \, \mu$.
In both cases, the velocity of the stirrer is above the steep onset of dissipation associated with the breakdown of superfluidity.
For the repulsive stirrer we use $v=3.7\, \mathrm{mm/s}$, for the attractive stirrer we use $v = 4.4\, \mathrm{mm/s}$.
The time evolution of the phase during stirring displays strong phase gradients for repulsive stirring, while the attractive stirrer only creates a smoothly varying phase field, associated with the creation of phonons.
The phase evolution for the repulsive stirrer is characterized by the creation of vortex--anti-vortex pairs. This can be confirmed by displaying   the phase winding around each plaquette, in particular $\sum_{\Box} \delta \phi(x,y) = \delta_x\phi(x,y) + \delta_y\phi(x+l,y)+\delta_x\phi(x+l,y+l)+\delta_y\phi(x,y+l)$, where the phase differences between sites is taken to be $\delta_{x/y} \phi(x,y)  \in (-\pi, \pi]$. 
We display the calculated phase winding in Fig. \ref{fig_vortices_rep}, where closely bound pairs of phase winding $+2\pi$ and $-2\pi$ are observed, corresponding to vortex pairs. 
To display the decay with clarity, we also sketch it in  Fig. \ref{fig_sketch_vortexpair}.
We note that this mechanism  of vortex-pair-induced dissipation has not been considered or suggested in the literature. However, a recent experimental study displays similarities, Ref. \cite{Shin2015}.

\begin{figure}[]
\includegraphics[width=0.88\linewidth]{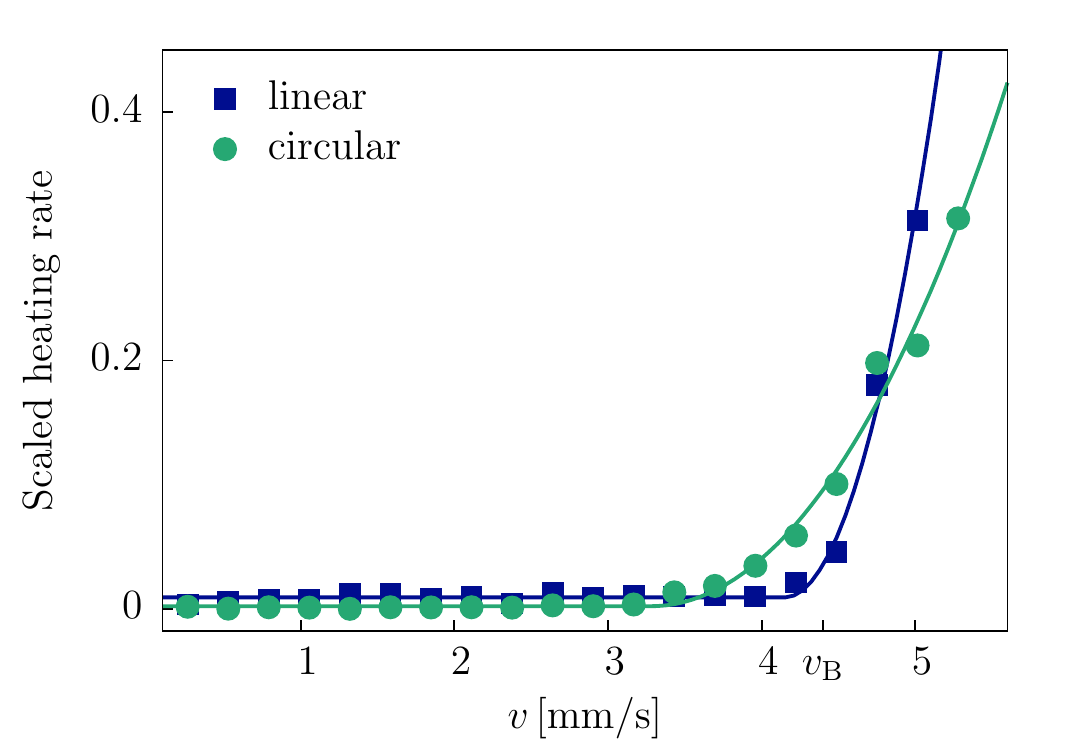}
\caption{(Color online) Simulated heating rates obtained for linear and circular stirring, with an attractive stirrer of strength $V_0=-0.03\, \mu$. The onset of heating occurs at a critical velocity $v_c=4.2$ and $3.3 \, \mathrm{mm/s}$ for the linear and circular stirring, respectively. Here, $v_c$ is determined using Eq. \ref{eq_fit_function}, and the fitted curves are shown by the solid lines. }
\label{fig_circ_lin}
\end{figure}
\begin{figure*}[]
\includegraphics[width=1.0\linewidth]{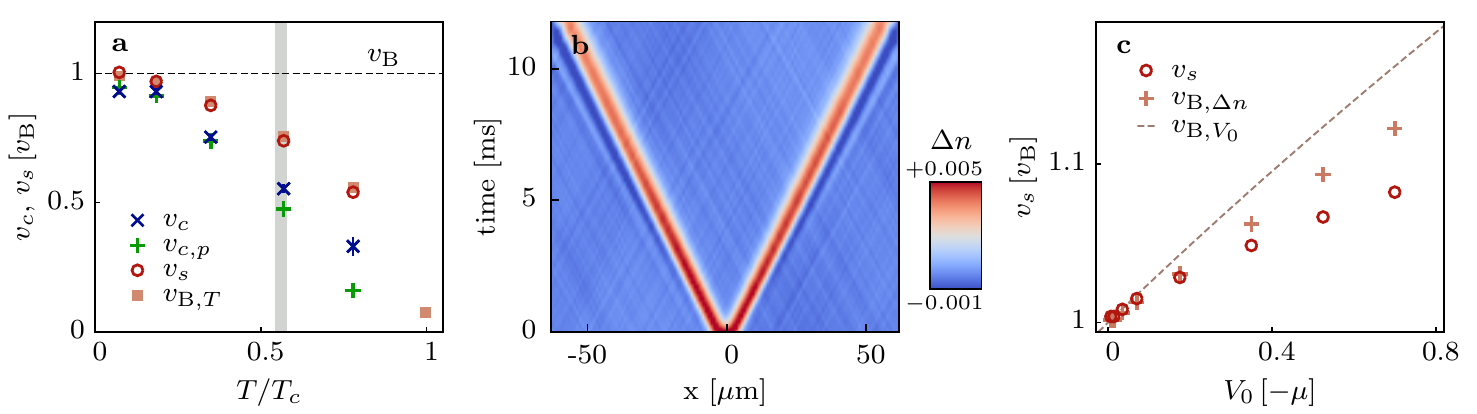}
\caption{(Color online) (a) 
The temperature dependent critical velocities $v_c$ (crosses) obtained from the simulation are compared to the critical velocities $v_{c, p}$ (plus symbols) obtained from the analytical heating rate, Eq. \ref{eg_heatingRate_damping2} using the fitting function in Eq. \ref{eq_fit_function}. They are in good agreement but significantly below the sound velocity obtained from the simulation $v_s$ (open circles) and from the condensate density at finite $T$, $v_{\mathrm{B}, T}$ (filled squares).  
The gray, vertical thick line corresponds to the mean-field energy $\mu \approx 0.56\, k_\mathrm{B}T_c$. 
(b) The propagation of the outgoing density waves after the perturbation is turned off at time $t=0$. 
(c) The sound velocity obtained in the simulation is influenced by the initial potential strength $V_0$. 
As is shown here for the various potential strengths $V_0$ used in the simulation, and is compared to the sound velocity obtained from the local density excess $v_{\mathrm{B}, \Delta n}$ (plus symbols) and from the applied potential $v_{\mathrm{B}, V_0}$ (dashed line).}
\label{fig_vc_temp}
\end{figure*}

\section{Circular versus linear stirring}\label{sec_circ_vs_lin}

In a trapped condensate, a circularly moving stirrer is a natural choice to probe the system at a fixed density.
 However, here we point out that the circular stirring motion results in an intrinsic reduction of the onset of dissipation to lower velocities, compared to a linear stirring motion of the same velocity.  
  In Fig. \ref{fig_circ_lin} we show a comparison of the heating rate due to linear and circular stirring motion.
 We calculate the heating rates of a homogeneous condensate at $T=0.07 \, T_c$. The circular motion of the stirrer has a radius $R_0$ and a frequency $\omega_m$, the velocity of the stirrer  is therefore given by $v=R_0 \omega_m$. In this example, we choose $R_0=10\, \mu \mathrm{m}$ and calculate the heating rates as a function of stirring velocity $v$. The onset of dissipation for the circular stirring motion is reduced, compared to the linear motion.

  
 We now estimate the heating rate analytically, within Bogoliubov theory, in analogy to the previous results for a linear stirring motion.
We consider a stirring potential of the form:
\begin{equation} \label{eq_potential_circ}
V({\bf r}, t) = V_0 \exp \Bigl(- \frac{ \bigl(x -x_s(t) \bigr)^2 + \bigl(y -y_s(t) \bigr)^2}{2\sigma^2} \Bigr),
\end{equation}
with $\bigl( x_s(t), \, y_s(t) \bigr) = R_0 \bigl( \cos(\omega_m t), \sin(\omega_m t) \bigr)$. As described in Appendix \ref{App_eRate_circ}, the equation of motion is now solved by
\begin{equation} \label{eq_ansatzcirc}
\hat{b}_{\bf k}(t) = e^{-i\omega_k t} \hat{b}_{\bf k} + A'_{\bf k}(t),
\end{equation}
with
\begin{align} \label{eq_Akprime_sol}
A^\prime_{\bf k}(t) &= -\frac{2i}{\hbar} (u_k + v_k) \sqrt{N_0} V_\bk \sum_{\nu=-\infty}^{\infty} i^\nu e^{-i\nu\phi}  J_\nu(kR_0)  \nonumber \\ 
& \quad \times \frac{\sin\bigl[(\omega_k -\nu\omega_m)t/2 \bigr] }{ (\omega_k -\nu\omega_m)} e^{-i(\omega_k -\nu\omega_m)t/2},
\end{align}
where $J_\nu(kR_0)$ are the Bessel functions of the first kind of order $\nu$. With this, and by integrating out the polar angle, the change of energy is
\begin{align} \label{eq_circ_energySecond}
& \langle \hat{\cH}_{0}(t)\rangle - \langle \hat{\cH}_{0}(0)\rangle = \frac{2A }{\pi \hbar} \int_0^\infty \mathrm{d}k \, k \omega_k  (u_k + v_k)^2 N_0 |V_\bk|^2   \nonumber \\ 
&  \qquad  \times \sum_{\nu} J_\nu^2(kR_0)  \frac{\sin^2\bigl[ (\omega_k -\nu\omega_m)t/2 \bigr] }{(\omega_{k} -\nu\omega_m)^2}.
\end{align}
This results in the heating rate
\begin{align} \label{eq_circ_heatingRateDelta}
\frac{dE}{dt} &= \frac{A }{\hbar} \sum_{\nu=-\infty}^{\infty} \int_0^\infty \mathrm{d}k \, k \omega_k (u_k + v_k)^2 N_0 |V_\bk|^2 J_\nu^2(kR_0) 
\nonumber \\ 
&  \quad  \times \delta(\omega_{k} -\nu\omega_m).
\end{align}
We approximate the sum over the Bessel function index $\nu$ by an integral over a continuous variable, and obtain
\begin{equation} \label{eq_circ_heatingRate}
\frac{dE}{dt} = \frac{A }{\hbar \omega_m} \int \mathrm{d}k \, k \omega_k  (u_k + v_k)^2 N_0 |V_\bk|^2 J_{\frac{\omega_k}{\omega_m}}^2(kR_0).
\end{equation}
To obtain the velocity scale of the onset of dissipation we note that for large indices 
$J_{\omega_k/\omega_m}^2(kR_0)$ has its maximum at \cite{Watson1966}
\begin{equation} \label{eq_Jsq_max}
k_mR_0 \approx \frac{\omega_k}{\omega_m}  + C_0 \Bigl( \frac{\omega_k}{\omega_m} \Bigr)^{1/3},
\end{equation}
where $C_0\approx 0.809$. Note that $kR_0=\omega_k/\omega_m$ is the same statement as $vk=\omega_k$, i.e. we recover approximately the Landau criterion. Eq. \ref{eq_Jsq_max} can also be written as 
\begin{equation} \label{eq_Jsq_max2}
vk_m \approx \omega_k \Bigl(1+  C_0 \Bigl( \frac{a_c}{R_0\omega_k^2} \Bigr)^{1/3}   \Bigr),
\end{equation}
where $a_c= v^2/R_0$ is the centripetal acceleration. The maximum thus approaches the location of the $\delta$-function in Eq. \ref{eg_heatingRate}, as $R_0$ is increased. The width of that maximum is $\sigma_B = 1/(2\sqrt{C_0})\times (\omega_k/\omega_m)^{1/3}$.
The ratio of the width to the location of the maximum is
\begin{equation} \label{eq_Jsq_max_location}
\frac{\sigma_B}{k_mR_0} = \frac{1}{2\sqrt{C_0}} \Bigl( \frac{a_c}{R_0\omega_{k}^2} \Bigr)^{1/3}.
\end{equation}
 This is the intrinsic broadening of the peak due to the accelerated nature of the circular motion.
 As we increase the radius $R_0$ of the circular motion, the ratio of the centripetal acceleration and the radius $a_c/R_0$ approaches zero. Thus, the function $J_{\omega_k/\omega_m}^2(kR_0)$ becomes more and more narrow around its maximum, and the sharp onset of dissipation for a linearly moving stirrer is recovered. 

 Using the expressions for $V_\bk$ and $(u_k + v_k)$ in Eq. \ref{eq_circ_heatingRate}, we obtain
\begin{equation} \label{eq_circ_heatingRate2}
R_{\mathrm{sc}} =  \frac{2 \pi^2 \sigma^4}{m A\omega_m}  \int_0^\infty \mathrm{d}k\, k^3  \exp(-k^2\sigma^2) J_{\omega_k/\omega_m}^2(kR_0). 
\end{equation}
We show this analytical prediction in Fig. \ref{fig_circ_lin_comp}(b) in comparison to heating rates that were obtained numerically, which gives 
good agreement. 
We also show the analytical prediction for the lattice system by the dotted line in Fig. \ref{fig_circ_lin_comp}(b), see Eq. \ref{Beq_eRateLattice} in Appendix \ref{App_eRate_circ}, giving very good agreement.

 Written in dimensionless form, in analogy to Eq. \ref{eq_aRate_norm2}, we rewrite Eq. \ref{eq_circ_heatingRate2} as 
\begin{equation} \label{heatingratecirc}
R_{\mathrm{n}}= \sqrt{8}\pi \tilde{v} \int_0^{\sigma k_u} \mathrm{d} \tilde{k}\, \tilde{k}^3  e^{-\tilde{k}^2} J_{\tilde{v} \tilde{k} \sqrt{\frac{\tilde{k}^2} {2}+ \frac{\sigma^2}{\xi^2}} }^2(\tilde{k}R_0/\sigma),
\end{equation}
where $\tilde{k}=k \sigma$, and $\tilde{v}= R_0 \xi v_\mathrm{B}/(v\sigma^2)$. We show the comparison of the heating rates $R_\mathrm{n}$ for the linear and circular stirring in Fig. \ref{fig_analytic_circ_lin}.
Both this figure and Fig. \ref{fig_circ_lin_comp} display the difference between the cases of  linear and circular stirring. 
For circular stirring, both the width and the location of the peak depend on the stirring radius $R_0$, described by Eq. \ref{eq_Jsq_max_location}. 
Furthermore, the heating rate has a weak oscillatory behavior for large $v$, which is due to the oscillatory behavior of the Bessel functions, controlled by the radius $R_0$. As $R_0$ is increased, the first peak builds up, and the oscillatory behavior shows a smaller and smaller period in $v$, visible in the inset of Fig. \ref{fig_circ_lin_comp}(b). For very large $R_0$ the onset of dissipation approaches that of  linear stirring.

\section{Influence of temperature on $v_c$} \label{sec_vc_temp}

%
\begin{figure}[]
\includegraphics[width=0.85\linewidth]{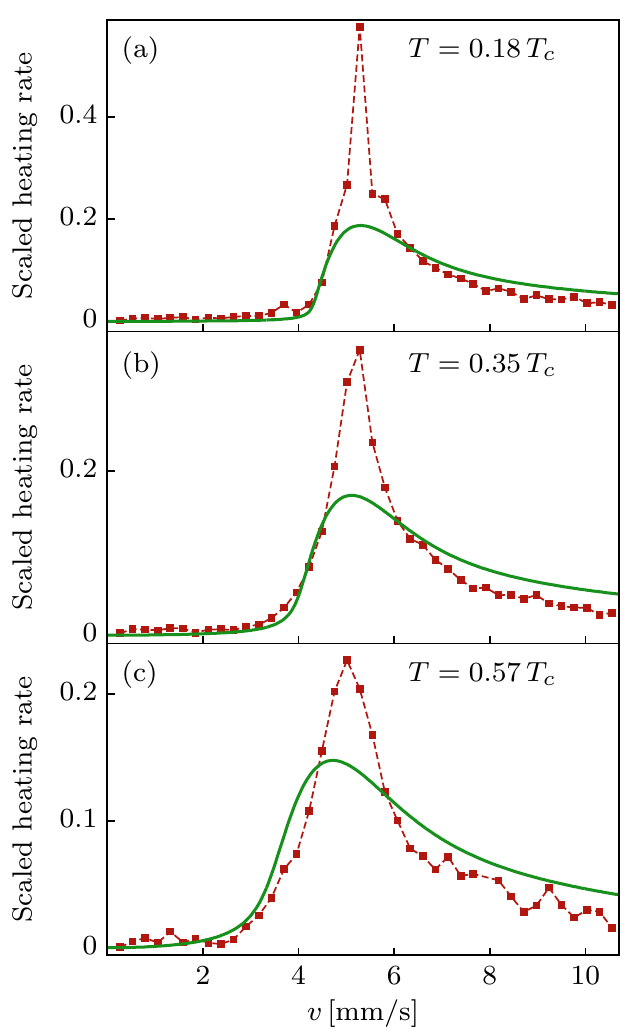}
\caption{(Color online) The simulated heating rate (red squares connected with a dashed line) is compared to the analytical prediction (green, continuous lines) for linear stirring at finite temperatures: (a) $T=0.18 \,T_c$, (b) $T=0.35 \, T_c$, and (c) $T=0.57 \,T_c$. The analytical results are obtained for a continuum system using Eq. \ref{eg_heatingRate_damping2}. }
\label{fig_lin_comp_temp}
\end{figure}

Up to now, we have considered temperatures that are small compared to the mean-field energy, and also small compared to the typical temperatures used in experiments, such as in Ref. \cite{Weimer2015}.

In Fig. \ref{fig_vc_temp}(a) we show the temperature dependence of the critical velocity for a homogeneous condensate that is stirred with a linear motion. We use a lattice with $128\times32\times16$ sites and calculate the heating rates at different temperatures $T$. From the heating rates we determine the critical velocity using the fitting function in Eq. \ref{eq_fit_function}. We use the linear stirrer with strength $V_0=-0.03\, \mu$. We compare the temperature dependence of $v_c$ to the Bogoliubov velocity $v_{\mathrm{B}} \equiv \sqrt{gn/m}$ in Fig. \ref{fig_vc_temp}(a), where $n$ is the total density. We compare $v_c$ to the Bogoliubov velocity $v_{\mathrm{B}, T} \equiv \sqrt{gn_{c}/m}$, where the numerically obtained $n_{c}$ is the condensate density at temperature $T$. At low temperatures $v_c$ and $v_{\mathrm{B},T}$ are close to each other, but start to deviate as the temperature is increased. As we demonstrate below, 
 this is due to the thermal broadening of the phonon modes.

In addition to the comparison to the estimated sound velocity we also determine the sound velocity $v_s$ numerically. 
  We use a Gaussian potential of width $\sigma$ along the $x$-axis at the center of the homogeneous system, with a strength $V_0=-0.03\, \mu$ that we slowly ramp up over $100\, \mathrm{ms}$ at the center and then abruptly turn it off.  This results in a density wave traveling outward, as shown in Fig. \ref{fig_vc_temp}(b).
  The density wave splits into two outgoing waves propagating along the $x$-axis. 
   From the linear propagation  we determine $v_s$.
  The example in Fig. \ref{fig_vc_temp}(b) is for $T=0.07\, T_c$. We  perform this simulation for a range of temperatures and show the temperature dependence in Fig. \ref{fig_vc_temp}(a). The results for $v_s$ are in excellent agreement with $v_{\mathrm{B},T}$. 
   This demonstrates that the reduction of the critical velocity is not entirely due to the thermal reduction of the sound velocity itself. We show below that the critical velocity can be recovered, if not only the reduced sound velocity of the phonon modes but also their thermal broadening is taken into account.
  
    We note that the velocities that are obtained in this way 
 are influenced by the initial potential strength $V_0$. This is shown in Fig. \ref{fig_vc_temp}(c), at $T=0.07\, T_c$, where we plot the velocity in units of $v_{\mathrm{B}}$ as a function of $V_0$. As the strength $V_0$ of the attractive potential is increased, the velocity $v_s$ increases. For comparison we calculate the Bogoliubov velocity $v_{\mathrm{B}, \Delta n}$ using the local density excess $\Delta n$ at the center owing to the applied potential $V_0$. In Fig. \ref{fig_vc_temp}(c) we show $v_{\mathrm{B}, V_0}/v_{\mathrm{B}} = \sqrt{1-|V_0|/(2\mu)}$ and $v_{\mathrm{B}, \Delta n}/v_{\mathrm{B}} =\sqrt{1+\Delta n/(2n)}$ by the dashed line and the plus symbols, respectively. Here, the factor $1/2$ appears because the excited wave packet is divided into two equal density waves. 
  For this reason, to obtain a reliable numerical estimate of the sound velocity, a sufficiently small value of the initial potential has to be used, as we did for the data depicted in    Fig. \ref{fig_vc_temp}(a).

 To expand on the analytical approach given above, we include non-zero thermal broadening of the phonon modes 
 due to Landau damping \cite{Hohenberg1965}. The $\delta$-distribution in the Bogoliubov estimate of the heating rate, Eq. \ref{eg_heatingRate}, is replaced by a Lorentzian distribution with the width given by \cite{Griffin2009}
\begin{equation} \label{eq_damping}
\Gamma_k = \frac{4}{3} \frac{a_sk_{\mathrm{B}}Tk}{\hbar}.
\end{equation}
$\Gamma_k$ is the width of a phonon mode with momentum $k$. The heating rate is then 
%
\begin{equation} \label{eg_heatingRate_damping}
\frac{dE}{dt} = \frac{2}{\hbar} \sum_{\bf k} \omega_k (u_k + v_k)^2 N_0|V_\bk|^2 \frac{\Gamma_k}{(\omega_k - {\bf vk})^2 + \Gamma_k^2}. 
\end{equation}
After substituting $(u_k + v_k)^2$ and $|V_\bk|^2$, we rewrite Eq. \ref{eg_heatingRate_damping} as
\begin{equation} \label{eg_heatingRate_damping2}
R_\mathrm{sc} = \frac{\sigma^4}{m A} \int \mathrm{d}^2k \, k^2 e^{-k^2\sigma^2} \frac{\Gamma_k}{(\omega_k - vk_x)^2 + \Gamma_k^2},
\end{equation}
where $k^2=k_x^2+k_y^2$. We numerically evaluate Eq. \ref{eg_heatingRate_damping2} for a continuum system and compare this prediction with the simulated heating rate at different temperatures $T$ in Fig. \ref{fig_lin_comp_temp}. The analytical results are in quantitative agreement with the simulations. To determine $v_c$ we fit the analytical heating rate with Eq. \ref{eq_fit_function}. We show the temperature dependence of the analytical $v_c$ in Fig. \ref{fig_vc_temp}(a). For temperatures $T \leq \mu/k_\mathrm{B}$ the predictions for $v_c$ are in agreement with the simulated $v_c$.

\section{Uniform versus trapped system} \label{sec_trapped_system}
\begin{figure}[]
\includegraphics[width=1.0\linewidth]{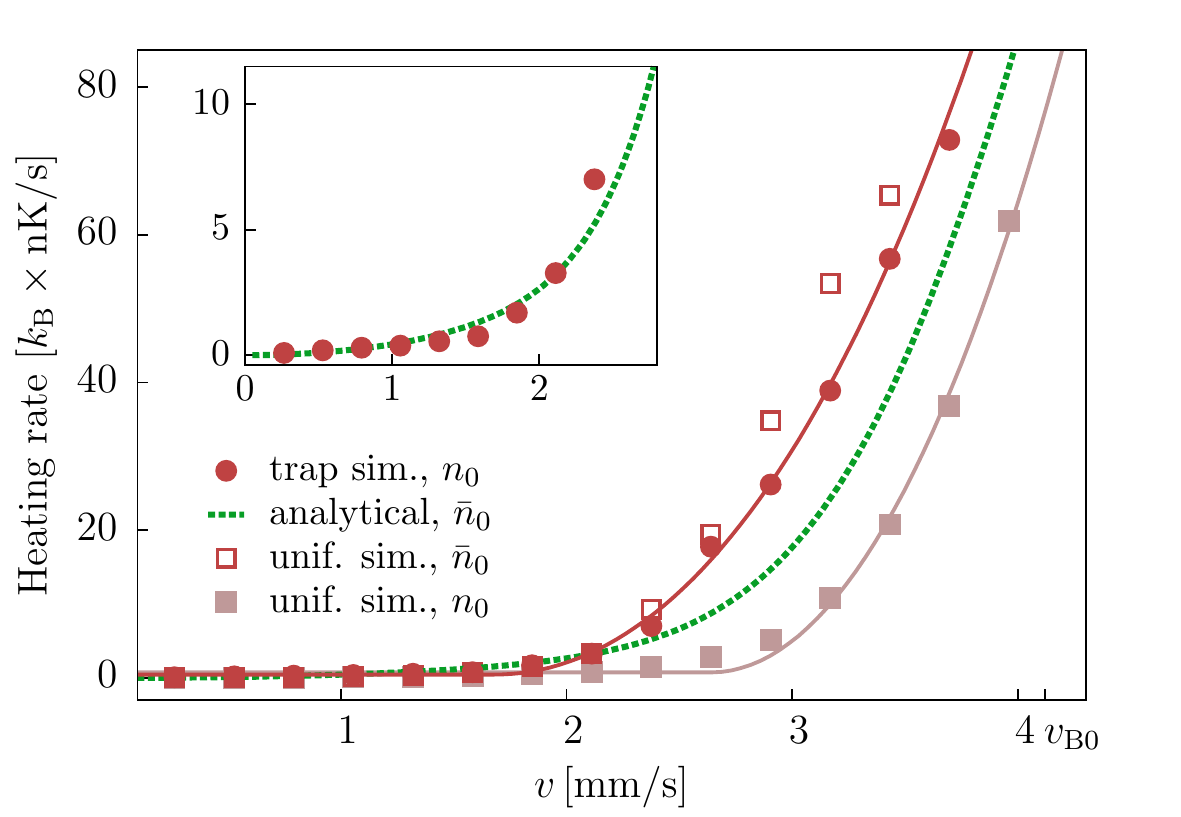}
\caption{(Color online) The numerically obtained heating rates (circles), for a trapped condensate, are compared to the analytical predictions (dotted line). The analytical results use the effective  condensate density obtained for a Thomas-Fermi profile. For comparison, we also show the numerical results for a homogeneous density of $0.42\, \mu\mathrm{m}^{-3}$ and for $0.28\, \mu\mathrm{m}^{-3}$, shown by the filled and empty squares, respectively. 
The density of $0.42\, \mu\mathrm{m}^{-3}$ corresponds to the center density at stirrer location, the density of $0.28\, \mu\mathrm{m}^{-3}$ corresponds to the effective density at stirrer location.
The inset shows the trapped and analytical results on a smaller scale, demonstrating good agreement at the onset of dissipation.}
\label{fig_realistic}
\end{figure}
%
    We now consider an inhomogeneous density distribution, as it is realized in an atomic trap, and the influence of the inhomogeneity on the heating rate. 
  As an example we show the numerically obtained heating rates for a trapped system  in Fig. \ref{fig_realistic}, for parameters chosen in accordance with the experiment \cite{Weimer2015}, corresponding to the experimental data point at $-1/k_Fa \approx -3.5$ in Fig. \ref{fig_vc_comp}.  The temperature is set to be $T=10\, \mathrm{nK}$. For comparison, we also simulate a uniform system with an average density of $0.42\, \mu\mathrm{m}^{-3}$, which is the density $n_0$ at the stirrer location in the trap, see \cite{Stirrer_density}. The onset of dissipation in the trapped system occurs at a smaller velocity as compared to the uniform case, see Fig. \ref{fig_realistic}. 
  

 
 This leads us to consider an effective, reduced density instead of the maximal density of the trapped system.
   We determine the effective density $\bar{n}_{0}$ by averaging over a Thomas-Fermi distribution, \cite{Stringari1998}, and obtain an effective density of $2/3$ times the central density. 
     We now use this effective density in a simulation of a homogeneous system. We show the results for the uniform case with the effective density $\bar{n}_0$ at the stirrer location by the empty squares in Fig. \ref{fig_realistic}.
    The comparison between the simulation for the trapped system and homogeneous system with the effective density shows good agreement.
  
 
 
 
 Furthermore, we use this effective density in the analytical description of the heating process, in addition to 
  extending the heating rate for circular stirring, Eq. \ref{eq_circ_heatingRateDelta}, to finite temperatures by replacing the $\delta$-function with a Lorentzian distribution. Therefore, the finite temperature heating rate is
\begin{align} \label{eq_circ_heatingRate_damping}
\frac{dE}{dt} &= \frac{A }{\pi \hbar} \sum_{\nu=-\infty}^{\infty} \int_0^\infty \mathrm{d}k \, k \omega_k  (u_k + v_k)^2 N_0|V_k|^2 J_\nu^2(kR) \nonumber \\ 
& \quad  \times \frac{\Gamma_k }{(\omega_k-\nu \omega_m)^2 + \Gamma_k^2},
\end{align}
where the Landau damping $\Gamma_k$ is given by Eq. \ref{eq_damping}. Rewriting Eq. \ref{eq_circ_heatingRate_damping} as
\begin{align} \label{eq_circ_heatingRate_damping2}
R_{\mathrm{sc}} &= \frac{2 \pi \sigma^4 }{m A} \sum_{\nu} \int_0^\infty \mathrm{d}k \,  \frac{k^3\exp(-k^2 \sigma^2)  J_\nu^2(kR) \Gamma_k }{(\omega_k-\nu \omega_m)^2 + \Gamma_k^2}.
\end{align}
We evaluate Eq. \ref{eq_circ_heatingRate_damping2} using the effective density $\bar{n}_0$ and the temperature $T=10\, \mathrm{nK}$. We show this analytical prediction by the dotted line in Fig. \ref{fig_realistic}. This indeed captures the onset of dissipation well.

\section{Comparison to experiment} \label{sec_comparison}
\begin{figure}[]
\includegraphics[width=1.0\linewidth]{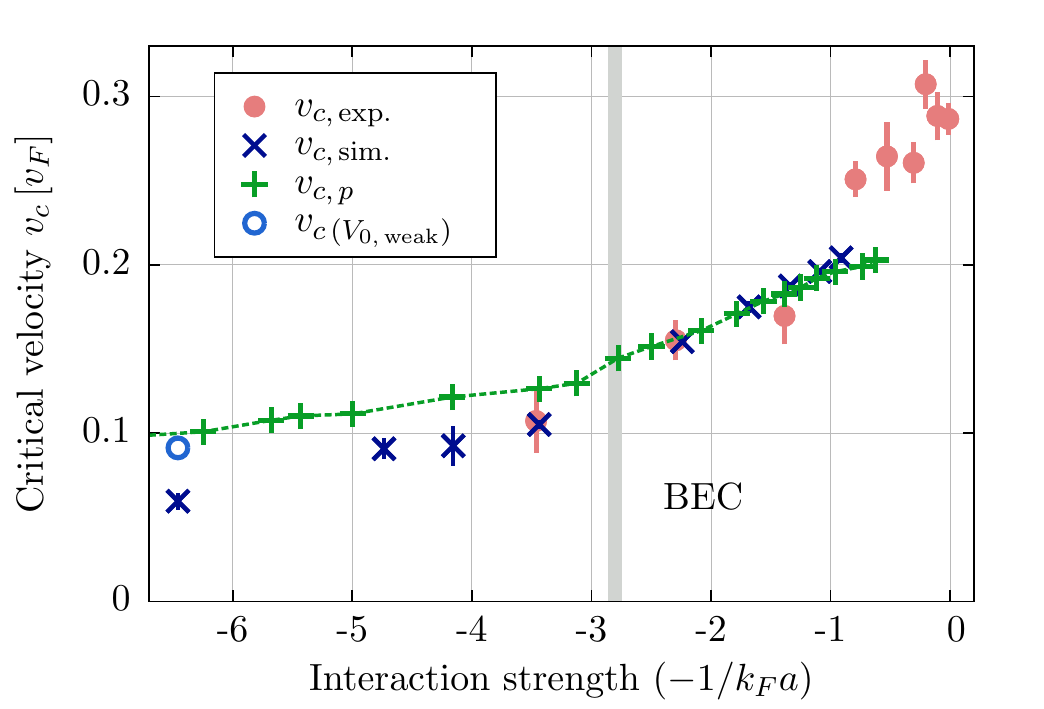}
\caption{(Color online) The experimentally measured (filled circles), the numerically simulated (crosses) and the analytical prediction (plus symbols)  for the critical velocity $v_c$ in units of the Fermi velocity $v_F$ in the BEC regime. The numerical results and the experimental results show quantitative agreement throughout the BEC regime, excluding the strongly correlated regime. The analytical prediction shows good agreement for the weak stirrer regime.
 In addition, we show the numerically obtained $v_c$ for a weak stirrer of strength $V_0=-h \times 2.8\, k\mathrm{Hz}$, shown by the empty circle. This recovers the analytical prediction, which assumes a weak stirrer. }
\label{fig_vc_comp}
\end{figure}
Finally, we compare the simulation and the analytical results to the experimental results \cite{Weimer2015} in the BEC regime.  
 We use the same parameters as the experiment and determine the critical velocities $v_c$ via the fitting method described before, for various scattering lengths $a_s \equiv a_{DD}$, $a_{DD}$ being the dimer-dimer scattering length.  
  The dimer-dimer scattering length is calculated from the atom-atom scattering length $a$ using the relation $a_{DD}=0.6a$ \cite{Petrov2004, Astrakharchik2004}.
  The experimentally measured and the simulated critical velocities $v_c$ in units of the Fermi velocity $v_F$ are plotted against the interaction strength in Fig. \ref{fig_vc_comp}. The Fermi velocity of a non-interacting gas is given by $v_F =(2\hbar/m_a)^{1/2} \times (\omega_r^2 \omega_z 6N)^{1/6}$. $m_a$ is the mass of a $^6$Li atom, $N$ is the total number of atoms, and $\omega_r$ ($\omega_z$) is the trap frequency in the radial (transverse) direction. The interaction strength is shown in terms of the dimensionless quantity $-1/k_Fa$, where $a$ is the atom-atom scattering length and the Fermi wavevector $k_F=m_av_F/\hbar$.  
         Throughout the BEC regime the experimentally measured and the numerically obtained $v_c$ are in good agreement. 
  We note that the simulations start to deviate from the experimental results  in the strongly interacting limit $-1/k_Fa\gtrsim -1$, where the simulation method becomes unreliable due to strong correlations. 
   Furthermore, we display the analytical prediction based on the finite-temperature heating rate, Eq. \ref{eq_circ_heatingRate_damping2}, for a circularly moving stirring potential, and for the effective density based on the Thomas-Fermi approximation. From this heating rate we determine an estimate of the critical velocity, again via fitting.

  Again, we find remarkable agreement for interactions $-1 > -1/k_Fa \gtrsim -2.8$. The theoretical prediction starts to deviate  both for the strongly-correlated regime, and for $-1/k_Fa < -2.8$. While the breakdown for the strongly-correlated regime is again due to the inapplicability of a weak-coupling method, the disagreement  for weak interactions is due to entering the strong-stirrer regime, i.e. $|V_0| >\mu_0$. 
$\mu_0$ is the mean-field energy at the stirrer location.
     The regime where the stirrer strength is comparable to the mean-field energy is shown by the gray, vertical thick line in Fig. \ref{fig_vc_comp};  to the left of this line we are in the strong-stirrer limit. 
      We demonstrate that this is the correct interpretation, by showing a weak stirrer given by $V_0=-h \times 2.8\, k\mathrm{Hz}$, in contrast to the experimentally used $V_0=-h \times 8.5\, k\mathrm{Hz}$,  by the empty circle in  Fig. \ref{fig_vc_comp}. This simulation indeed reproduces the analytical prediction.

\section{conclusions} \label{sec_conclusion}
\begin{figure}[]
\includegraphics[width=1.0\linewidth]{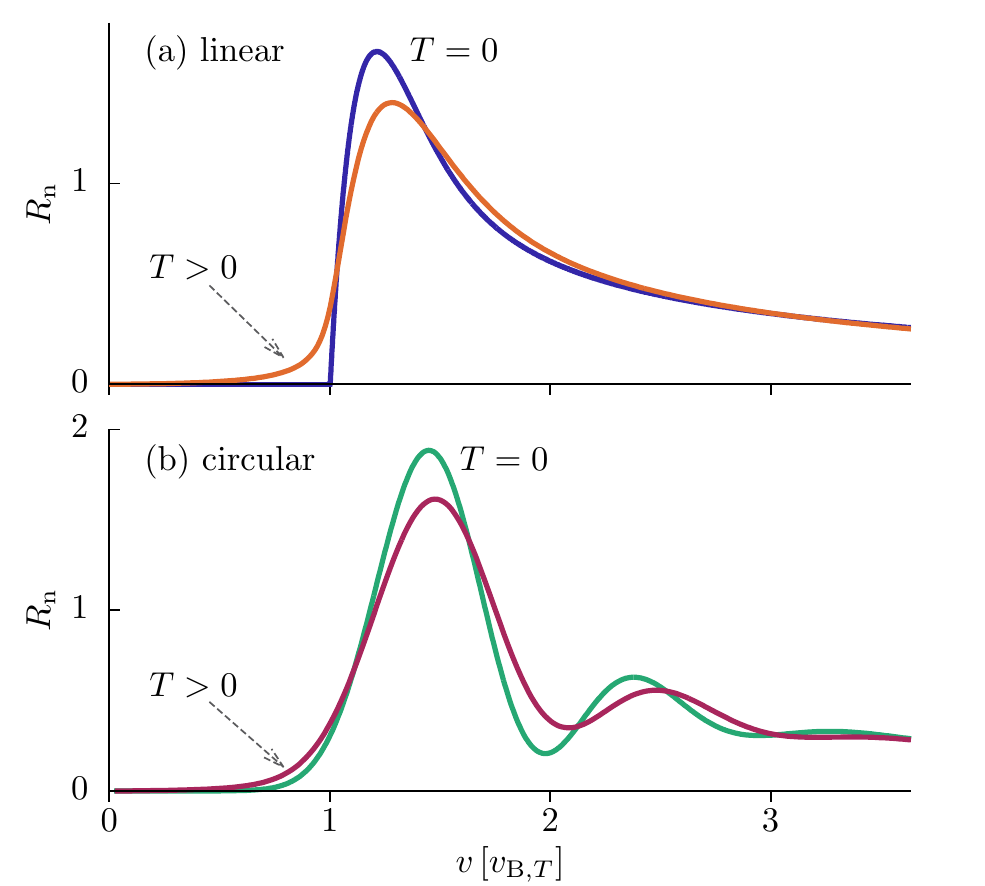}
\caption{(Color online) Analytical heating rates $R_\mathrm{n}$ as a function of stirring velocity $v$ at temperature $T=0$ and $10\, \mathrm{nK}$. Panel (a) corresponds to linear stirring, whereas panel (b) shows the results for circular stirring with radius $R_0=10\, \mu \mathrm{m}$.}
\label{fig_eRate_summary}
\end{figure}
In conclusion, we have demonstrated a quantitative understanding of the stability of superfluidity of  a Bose-Einstein condensate stirred with an attractive potential. 
  While we show that a similar, repulsive perturbation results in a decay mechanism due to vortex--anti-vortex pairs, which had not been considered before, we find that the choice of an attractive perturbation results in phononic decay only, while the influence of vortices is suppressed.
   We then continue to identify the reason why,  despite this phonon-only mechanism, the critical velocity that was obtained in the experiment  \cite{Weimer2015} is still noticeably smaller than the phonon velocity itself.
The contributing factors are the circular motion of the stirrer, which results in an intrinsic broadening due to an accelerated motion, and the thermal broadening. With these factors taken properly into account, a quantitative, even analytical, agreement is achieved. 
 The analytical prediction for the heating rate as a function of the stirring velocity is summarized in Fig. \ref{fig_eRate_summary}. While the
  zero-temperature, linear stirring process gives the phonon velocity as the critical velocity, the onset of dissipation is noticeably reduced due to the thermal broadening and using an accelerated stirrer. 
 Beyond giving a quantitative understanding of the experiment reported in Ref.  \cite{Weimer2015}, this study thus provides a general blueprint to analyze the breakdown of superfluidity, disturbed with a weak, localized pertubation, in a typical experimental setting.

\section*{Acknowledgement}
  We acknowledge support from the Deutsche Forschungsgemeinschaft through the SFB 925, GRK 1355, and the Hamburg Centre for Ultrafast Imaging, and from the Landesexzellenzinitiative Hamburg, supported by the Joachim Herz Stiftung.

\appendix

\section{Simulated heating rate}\label{App_sim_eRate}
\begin{figure}[]
\includegraphics[width=1.0\linewidth]{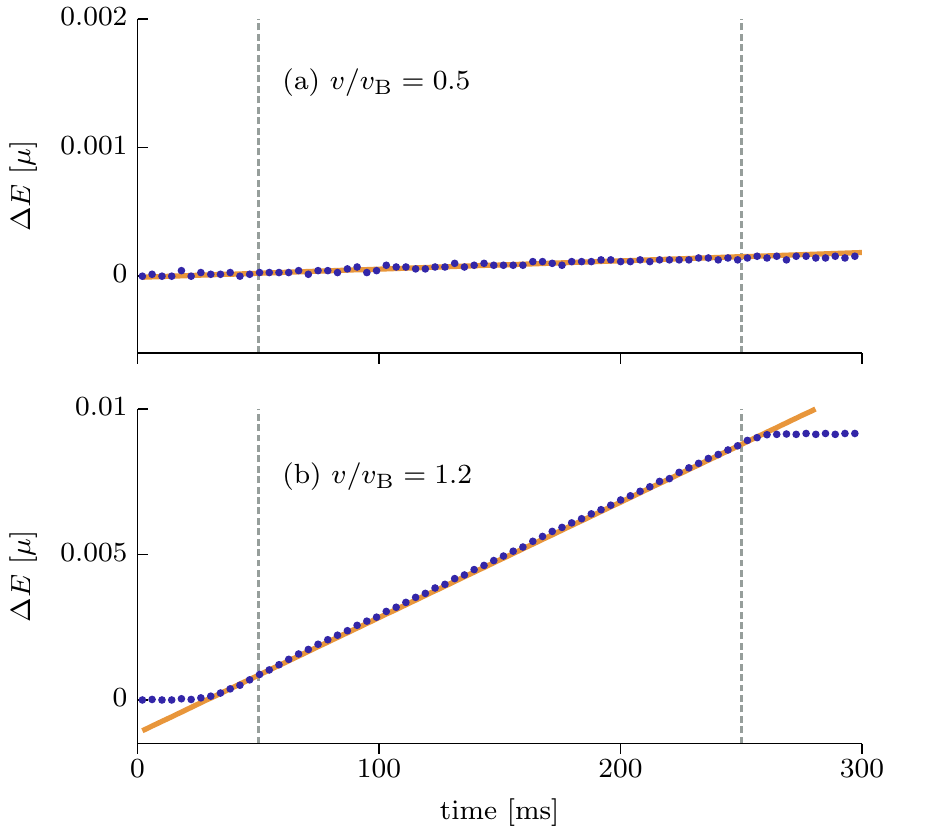}
\caption{(Color online) The total energy change $\Delta E$ (blue circles) during stirring of a homogeneous system with a linear stirrer of strength $V_0=-0.03\, \mu$. The simulated energy is normalized by the total number of atoms. Panel (a) corresponds to a stirring velocity $v<v_\mathrm{B}$, whereas panel (b) shows the total energy change for $v>v_\mathrm{B}$. We fit the linear energy increase during stirring with a linear function. The fitting duration is marked by the two vertical, dashed lines, and the fitted curves are shown by the continuous lines. }
\label{fig_energy_stir}
\end{figure}

In this section we illustrate how we obtain the heating rate from the numerical simulation method described in Sec. \ref{sec_sim_method}. To perform the simulation we discretize the real space and represent the continuous Hamiltonian, Eq. \ref{eq_hamil}, by the discrete Bose-Hubbard Hamiltonian \cite{Jaksch1998} on a three-dimensional square lattice:
\begin{equation} \label{eq_hubbard}
H_0 = -J \sum_{\langle i j \rangle} (\psi_i^\ast \psi_j + \psi_j^\ast \psi_i) + \frac{U}{2} \sum_i n_i^2
 + \sum_i V_i n_i,
\end{equation}
where $\psi_i$ and $n_i=|\psi_i|^2$ are the complex-valued field and the density at site $i$, respectively; $\langle ij \rangle$ represents nearest-neighbor bonds. For a lattice discretization length $l$, the Bose-Hubbard parameters are related to the continuum parameters, cp. Ref. \cite{Mora2003}, by $J=\hbar^2/(2ml^2)$ and $U=gl^{-3}$. $m$ is the atomic mass and $g=4\pi a_s \hbar^2/m$ is the interaction parameter; $a_s$ is the $s$-wave scattering length. The external potential $V_i$ describes the harmonic trap, $V_{\mathrm{trap},i}=m(\omega_r^2r^2+\omega_z^2z^2)/2$. 
$\omega_r$ ($\omega_z$) is the trap frequency in the radial (transverse) direction, and $r=(x^2+y^2)^{1/2}$ is the radial coordinate. 

 As an example we choose the homogeneous system described in Sec. \ref{sec_heating_rate} with linear stirring. We initialize the system in a thermal state at temperature $T$. The initial states are generated from a grand canonical ensemble via a classical Metropolis algorithm. We then introduce the stirring potential described by Eq. \ref{stirringpot} using classical equations of motion. After following the stirring sequence described in Sec. \ref{sec_sim_method} we calculate a thermal ensemble of the total energy $E = \langle H_0 \rangle$ using the unperturbed Hamiltonian in Eq. \ref{eq_hubbard} as a function of the stirring velocity $v$. We show two such cases, below and above the Bogoliubov velocity $v_\mathrm{B}$, for the linear stirring in Fig. \ref{fig_energy_stir}. During the stirring the energy increases linearly. From the linear energy increase we determine the induced heating by the stirrer.

\section{Analytical heating rate}
 We elaborate on calculations of the heating rate within the Bogoliubov approximation for both linear and circular stirring.
In Sec. \ref{App_BogHam} we describe the Bogoliubov Hamiltonian at temperature $T=0$.
In Sec. \ref{App_eRate_lin} we derive the heating rate for linear stirring, Eqs. \ref{eg_heatingRate} and \ref{eq_heatingRateBog}.
In Sec. \ref{App_eRate_circ} we show the full calculations of the heating rate for circular stirring, Eqs. \ref{eq_circ_heatingRateDelta} and \ref{eq_circ_heatingRate}.

\subsection{Bogoliubov Hamiltonian} \label{App_BogHam}
 In this section we describe the Bogoliubov Hamiltonian, the unperturbed Hamiltonian at temperature $T=0$, which is the starting point of our calculations. 
 For a uniform gas of $N$ interacting bosons in a box of volume $V$, the Hamiltonian in momentum space is given as
\begin{equation} \label{Aeq_Ham_kspace}
\hat{ \mathcal{H}_0}= \sum_{\bf k} \epsilon_k \hat{a}_{\bf k}^\dagger \hat{a}_{\bf k} + \frac{g}{2V}\sum_{\bf k, q, p} \hat{a}_{\bf k+p}^\dagger \hat{a}_{\bf q-p}^\dagger \hat{a}_{\bf q} \hat{a}_{\bf k},
\end{equation} 
where $\hat{a}_{\bf k}$ ($\hat{a}_{\bf k}^\dagger$) is the annihilation (creation) operator and $\epsilon_k=\hbar^2k^2/2m$. We diagonalize the Hamiltonian in Eq. \ref{Aeq_Ham_kspace} using the Bogoliubov transformation of the form $\hat{b}_{\bf k} = u_k \hat{a}_{\bf k}-v_k \hat{a}_{-{\bf k}}^\dagger$, where the Bogoliubov functions are given by $u_k^2= (\hbar \omega_k + \epsilon_k + gn_c)/(2\hbar \omega_k)$ and  $v_k^2= (- \hbar \omega_k + \epsilon_k + gn_c)/(2\hbar \omega_k)$, respectively. The diagonalized Hamiltonian is therefore  
\begin{equation} \label{Aeq_Ham_Bogliubov}
\hat{ \mathcal{H}_0}= \hbar \omega_0 + \sum_{{\bf k} \neq 0 } \hbar \omega_k \hat{b}_{\bf k}^\dagger \hat{b}_{\bf k}.
\end{equation} 
The dispersion of the quasiparticles is given by
\begin{equation}\label{Aeq_Bog_dispersion}
\hbar \omega_k= \sqrt{ \epsilon_k(\epsilon_k +2 m v_{\mathrm{B}}^2 ) },
\end{equation}
where $v_{\mathrm{B}} \equiv \sqrt{gn_c/m}$ is the sound velocity.

\subsection{Linear stirring } \label{App_eRate_lin}
Next, we introduce the stirring potential as a time-dependent perturbation. The total Hamiltonian of the system is then
\begin{equation}
\hat{ \mathcal{H}}(t)= \hat{ \mathcal{H}}_0 + \hat{ \mathcal{H}}_{s}(t),
\end{equation} 
where the time-dependent Hamiltonian $\hat{ \mathcal{H}}_{s}$ is given by 
\begin{equation}\label{Aeq_stir_Hamil}
\hat{ \mathcal{H}}_{s}(t) = \int \mathrm{d}{\bf r} \, V({\bf r},t) \hat{n}({\bf r}).
\end{equation} 
$\hat{n}({\bf r})$ is the density operator and $V({\bf r},t)$ is the stirring potential. We model the stirrer as a Gaussian with a width $\sigma$ and a strength $V_0$. For a linear stirring motion the Gaussian stirrer moves with a velocity ${\bf v} = (v_x, v_y)$. The stirring potential is therefore   
\begin{equation} \label{Aeq_stirrer}
V({\bf r},t) =  V_0\exp\Bigl(- \frac{(x -v_xt)^2+(y -v_yt)^2}{2\sigma^2} \Bigr).
\end{equation}
We Fourier transform the stirring potential using
\begin{equation}
V_\bk(t) = \frac{1}{V}\int \mathrm{d}{\bf r}\, e^{i{\bf kr}} V({\bf r},t).  
\end{equation}
Substituting $V({\bf r},t)$ using Eq. \ref{Aeq_stirrer} and making use of the property $L \delta_k = \int \mathrm{d}r \exp(-i kr)$, with $L$ being the system length, we obtain 
\begin{equation} \label{eq_FourierVs}
V_\bk(t) = \frac{2\pi V_0 \sigma^2}{A} \delta_{k_z} e^{i(k_xv_x +k_yv_y)t}  e^{-k_{x,y}^2\sigma^2/2},  
\end{equation}
where $A$ is the system area and the Kronecker delta $\delta_{k_z}$  is $1$ for $k_z=0$ and zero otherwise. Rewriting Eq. \ref{eq_FourierVs} as
\begin{equation} \label{Aeq_FourierVstir}
V_\bk(t)=\frac{2\pi V_0 \sigma^2}{A} \delta_{k_z} e^{i{\bf vk}t} e^{-k^2\sigma^2/2}.  
\end{equation}
We now express the density operator $\hat{n}({\bf r})$ in its Fourier representation, i.e. $\hat{n}_{\bf k} = \sum_{\bf k^\prime} \hat{a}_{\bf k^\prime}^\dagger \hat{a}_{\bf k^\prime+k}$. Assuming a macroscopic occupation of the mode ${\bf k}=0$, we expand 
\begin{equation}
\hat{n}_{\bf k} = \hat{a}_{-{\bf k}}^\dagger \hat{a}_0 +\hat{a}_0^\dagger \hat{a}_{\bf k}   + \sum_{ {\bf k^\prime}\neq 0} \hat{a}_{\bf k^\prime}^\dagger \hat{a}_{\bf k^\prime+k}. 
\end{equation} 
We linearize $\hat{n}_{\bf k}$ for small ${\bf k}$ using the Bogoliubov theory, where $\hat{a}_0$ and $\hat{a}_0^\dagger$ are replaced by $\sqrt{N_0}$. $N_0$ is the number of condensed atoms in the mode ${\bf k}=0$. Applying the Bogoliubov transformation, we obtain 
\begin{equation} \label{Aeq_Bogdensity}
\hat{n}_{\bf k} \approx \sqrt{N_0}(u_k + v_k) (\hat{b}_{-{\bf k}}^\dagger + \hat{b}_{\bf k} ). 
\end{equation} 
Now, we rewrite the stirring potential and density in its Fourier representation in the stirring Hamiltonian as
\begin{align}
\hat{ \mathcal{H}}_{s} &= \int \mathrm{d}{\bf r} \sum_{\bf k} e^{-i{\bf kr}} V_\bk(t) \frac{1}{V} \sum_{ {\bf k^\prime}} e^{-i{\bf k^\prime r}} \hat{n}_{\bf k^\prime}  \nonumber \\
&= \frac{1}{V} \sum_{{\bf k}, {\bf k^\prime}} V_\bk(t) \hat{n}_{\bf k^\prime} \int \mathrm{d}{\bf r} \exp\bigl(-i ({\bf k}+{\bf k}^\prime){\bf r} \bigr).
\end{align}
Using Eqs. \ref{Aeq_FourierVstir} and \ref{Aeq_Bogdensity}, and the property $V \delta_{{\bf k},-{\bf k}^\prime} = \int \mathrm{d}{\bf r} \exp\bigl(-i ({\bf k}+{\bf k}^\prime){\bf r} \bigr)$, we get
\begin{equation}
\hat{ \mathcal{H}}_{s} = \sum_{\bf k} V_\bk(t) \hat{n}_{\bf k}  = \sum_{\bf k} S_k (\hat{b}_{-{\bf k}}^\dagger + \hat{b}_{{\bf k}} ).
\end{equation}    
Here, $S_k \equiv 2\pi V_0 \sqrt{N_0} (u_{\bf k} + v_{\bf k})\sigma^2/A \times \delta_{k_z}e^{i{\bf vk}t} e^{-k^2\sigma^2/2}$.
The equation of motion for $\hat{b}_{\bf k} (t)$ is written as
\begin{equation} \label{Aeq_motion}
i \hbar \mathrm{d}_t \hat{b}_{\bf k}(t)  = \bigl[\hat{b}_{\bf k}(t), \hat{ \mathcal{H}_0} \bigr] + \bigl[\hat{b}_{\bf k}(t), \hat{ \mathcal{H}_s} \bigr].
\end{equation}  
We solve Eq. \ref{Aeq_motion} using the following ansatz: 
\begin{equation} \label{Aeq_ansatz}
\hat{b}_{\bf k}(t) = e^{-i\omega_k t} \hat{b}_{\bf k} + A_{\bf k}(t).
\end{equation}
Using Eq. \ref{Aeq_ansatz} the terms on the right-hand side of Eq. \ref{Aeq_motion} are 
\begin{align} \label{Aeq_motion_rhs1}
 & \bigl[\hat{b}_{\bf k}(t), \hat{ \mathcal{H}_0} \bigr] = \bigl[e^{-i\omega_kt} \hat{b}_{\bf k} + A_{\bf k}(t), \hbar \omega_0 + \sum_{ {\bf k^\prime}\neq 0} \hbar \omega_{k^\prime} \hat{b}_{\bf k^\prime}^\dagger \hat{b}_{\bf k^\prime}   \bigr] \nonumber \\
&= e^{-i\omega_k t} \sum_{ {\bf k^\prime}\neq 0} \hbar \omega_{k^\prime} \bigl[\hat{b}_{\bf k}, \hat{b}_{\bf k^\prime}^\dagger \hat{b}_{\bf k^\prime} ] =   e^{-i\omega_k t} \hbar \omega_k \hat{b}_{\bf k},
\end{align} 
and
\begin{align} \label{Aeq_motion_rhs2}
&\bigl[\hat{b}_{\bf k}(t), \hat{ \mathcal{H}_s} \bigr] = \bigl[e^{-i\omega_k t} \hat{b}_{\bf k} + A_{\bf k}(t),  \sum_{\bf k^\prime} S_{k^\prime} (\hat{b}_{-{\bf k}^\prime}^\dagger + \hat{b}_{{\bf k}^\prime} )   \bigr] \nonumber \\
&= e^{-i\omega_k t} \sum_{{\bf k}^\prime} S_{k^\prime} \bigl[\hat{b}_{\bf k}, \hat{b}_{-{\bf k}^\prime}^\dagger] = e^{-i\omega_k t} S_k.
\end{align}
Inserting Eqs. \ref{Aeq_motion_rhs1} and \ref{Aeq_motion_rhs2} into Eq. \ref{Aeq_motion}, which gives
\begin{equation} \label{Aeq_motionF}
i \hbar \mathrm{d}_t \hat{b}_{\bf k}(t)  = e^{-i\omega_k t} \hbar \omega_k \hat{b}_{\bf k} + e^{-i\omega_k t} S_k.
\end{equation} 
The time derivative of Eq. \ref{Aeq_ansatz} is
\begin{equation} \label{Aeq_ansatz_motion}
i \hbar \mathrm{d}_t \hat{b}_{\bf k}(t)  = e^{-i\omega_k t} \hbar \omega_k \hat{b}_{\bf k} + i \hbar \mathrm{d}_t A_{\bf k}(t),
\end{equation} 
which we equate with Eq. \ref{Aeq_motionF} and get
\begin{align} \label{Aeq_Ak(t)_motion}
\mathrm{d}_t A_{\bf k}(t) &= -\frac{i}{\hbar} e^{-i\omega_k t} S_k \nonumber\\
 &= -\frac{i}{\hbar} e^{-i(\omega_k - {\bf vk})t} (u_k + v_k)\sqrt{N_0} V_\bk. 
\end{align}
Here, $V_\bk = 2\pi V_0 \sigma^2/A \times \delta_{k_z} e^{-k^2\sigma^2/2}$. With the initial condition that the perturbation $V(t<0)=0$, for a finite stirring time $t$ we obtain
\begin{align} \label{Aeq_Ak(t)}
A_{\bf k}(t) &= \frac{-2i}{\hbar} (u_k + v_k)\sqrt{N_0} V_\bk \frac{\sin\bigl[ (\omega_k - {\bf vk})t/2 \bigr] }{(\omega_k - {\bf vk})} \nonumber \\
& \quad \times e^{-i(\omega_k - {\bf vk})t/2}.
\end{align}

  We now calculate the energy of the system at time $t$ using the dynamical evolution of the Bogoliubov operator created by Eqs. \ref{Aeq_Ham_Bogliubov} and \ref{Aeq_stir_Hamil}. The expectation value of the energy is 
\begin{equation} \label{Aeq_energyExpectationT}
\langle E(t)\rangle = \sum_{\bf k} \hbar\omega_k \langle \hat{b}_{\bf k}^\dagger(t) \hat{b}_{\bf k}(t)\rangle.
\end{equation}
Using the evolution of $\hat{b}_{\bf k}(t)$ given in Eq. \ref{Aeq_ansatz}, gives 
\begin{align} \label{Aeq_energyTime}
\langle E(t)\rangle &= \sum_{\bf k} \hbar\omega_k  \bigl(\langle\hat{b}_{\bf k}^\dagger\hat{b}_{\bf k}\rangle  + A_{\bf k}(t) \langle \hat{b}_{\bf k}^\dagger \rangle e^{i\omega_kt} \nonumber \\ 
& \quad + A_{\bf k}^*(t) \langle \hat{b}_{\bf k} \rangle e^{-i\omega_kt}  + |A_{\bf k}(t)|^2 \bigr).
\end{align}
The first-order terms in Eq. \ref{Aeq_energyTime} are zero, because $\langle\hat{b}_{\bf k}\rangle = \langle\hat{b}_{\bf k}^\dagger\rangle=0$. Therefore, the energy change due to the second-order term in perturbation is 
\begin{equation} \label{Aeq_energy_perturbation}
\langle\Delta E(t)\rangle = \sum_{\bf k} \hbar\omega_k  |A_{\bf k}(t)|^2.
\end{equation}
Substituting $|A_{\bf k}(t)|^2$ using Eq. \ref{Aeq_Ak(t)}, which gives
\begin{align} \label{Aeq_energySecond}
\langle\Delta E(t)\rangle &= \frac{4}{\hbar^2} \sum_{\bf k} \hbar\omega_k  (u_k + v_k)^2 N_0 |V_\bk|^2  \nonumber \\
& \quad \times \frac{\sin^2\bigl[ (\omega_k - {\bf vk})t/2 \bigr] }{(\omega_k - {\bf vk})^2}. 
\end{align}
In the $t\rightarrow \infty$ limit we use
\begin{equation} \label{Aeq_sincProp}
\lim_{t\rightarrow \infty} \frac{\sin^2(\alpha t/2)}{\alpha^2} = \frac{\pi}{2}t \delta(\alpha).
\end{equation}
Using Eq. \ref{Aeq_sincProp}, the energy in Eq. \ref{Aeq_energySecond} is proportional to $t$. Therefore, the heating rate is 
\begin{equation} \label{Aeg_heatingRate}
\frac{d E}{dt} = \frac{2\pi}{\hbar} \sum_{\bf k} \omega_k (u_k + v_k)^2 N_0 |V_\bk|^2  \delta(\omega_k - {\bf vk}).
\end{equation}
Using the explicit form of $|V_\bk|^2$ and $(u_k + v_k)^2 = \hbar k^2/(2m\omega_k)$, and $\sum_{k_z} \delta_{k_z}=1$ in Eq. \ref{Aeg_heatingRate}, we get
\begin{equation} \label{Aeg_heatingRate2}
 R_\mathrm{sc} = \frac{\pi \sigma^4}{m A} \int \mathrm{d}^2k \, k^2 \exp(-k^2\sigma^2) \delta(\omega_k - vk_x),
\end{equation}
where $k^2=k_x^2+k_y^2$, and $R_\mathrm{sc} \equiv (dE/dt)/(N_0V_0^2)$ is the rescaled heating rate. We have set the direction of the stirring velocity along the x-axis, i.e. $\bv = v \be_{x}$. In polar coordinates we integrate out the polar angle $\phi$. This we do by replacing the $\delta$-function using $\lim_{\epsilon \rightarrow0} \frac{\epsilon}{x^2+\epsilon^2}=\pi \delta(x)$. With this, 
\begin{equation} \label{Aeq_heatingRateBog}
 R_\mathrm{sc} =  \frac{2\pi \sigma^4}{m A}  \int_0^{k_u} \mathrm{d}k\,  \frac{k^3 \exp(-k^2\sigma^2)}{\sqrt{v^2k^2-\omega_k^2}}
\end{equation}
for $vk > \omega_k$, i.e. $v > v_{\mathrm{B}}$. The wavevector $k_u$ determines the maximum phonon momentum that can be created by a moving stirrer. Using the explicit form of $\omega_k$ and in the $k_u \rightarrow \infty$ limit, we rewrite Eq. \ref{Aeq_heatingRateBog} as
\begin{equation} \label{Aeq_heatingRateBog2}
 R_\mathrm{sc} =  \frac{2\pi \sigma^4}{m A}  \int_0^{\infty} \mathrm{d}k\,  \frac{k^2 \exp(-k^2\sigma^2)}{\sqrt{v^2-v_\mathrm{B}^2-\hbar^2 k^2/(2m)^2}}.
\end{equation}
Eq. \ref{Aeq_heatingRateBog2} can be solved analytically, giving 
\begin{align} \label{Aeg_heatingRateAnalytic}
R_\mathrm{sc}  &= \frac{\pi\sqrt{\pi}\sigma^2}{\hbar A} \nonumber \\
& \quad \times \Im \Bigl(-U \Bigl(1/2,0, 4m^2\sigma^2 (v_{\mathrm{B}}^2-v^2)/\hbar^2 \Bigr) \Bigr),
\end{align}
where $U(a,b,c)$ is confluent Hypergeometric function.

\subsection{Circular stirring} \label{App_eRate_circ}
In this section we derive the heating rate for a circular stirring motion with a radius $R_0$ and a frequency $\omega_m$. The stirring potential has the following form:
\begin{equation} \label{Aeq_potential_circ}
V({\bf r}, t) = V_0 \exp \Bigl(- \frac{ \bigl(x -x_s(t) \bigr)^2 + \bigl(y -y_s(t) \bigr)^2}{2\sigma^2} \Bigr),
\end{equation}
where $\bigl( x_s(t), \, y_s(t) \bigr)= R_0 \bigl( \cos(\omega_m t), \sin(\omega_m t) \bigr)$ are the locations of the Gaussian stirrer with width $\sigma$ and strength $V_0$. 
For a circular stirring motion, the Fourier transform of the potential is then
\begin{equation} \label{Aeq_potential_circFT}
V_\bk(t) = \frac{2\pi V_0 \sigma^2 }{A} \delta_{k_z} e^{i \bigl( k_x x_s(t) + k_y y_s(t) \bigr)} e^{-k^2\sigma^2/2}. 
\end{equation}
In polar coordinates we obtain
\begin{equation} \label{Aeq_FourierVstir_circ}
V_\bk(t) = \frac{2\pi V_0 \sigma^2}{A} \delta_{k_z} e^{ik R_0 \cos(\omega_m t-\phi)}  e^{-k^2\sigma^2/2}.
\end{equation}
We now solve the equation of motion described by Eq. \ref{Aeq_motion} using the stirring Hamiltonian defined by Eq. \ref{Aeq_stir_Hamil}, where we use the stirring potential for a circular stirring motion, Eq. \ref{Aeq_potential_circ}. The equation of motion is now solved by 
\begin{equation} \label{Aeq_ansatz_circ}
\hat{b}_{\bf k}(t) = e^{-i\omega_k t} \hat{b}_{\bf k} + A^\prime_{\bf k}(t).
\end{equation}
The solution of the equation of motion gives (similar to Eq. \ref{Aeq_Ak(t)_motion} for a linear stirring motion)
\begin{align} \label{Aeq_Ak(t)_prime}
\mathrm{d}_t A^\prime_{\bf k}(t) &= -\frac{i}{\hbar} e^{-i\omega_k t} S_k^\prime \nonumber \\
 & = -\frac{i}{\hbar} e^{-i\omega_k t} e^{ikR_0\cos(\omega_m t -\phi)} \nonumber \\
 & \quad \times (u_k + v_k)\sqrt{N_0} V_\bk. 
\end{align}  
Here, $S_k^\prime \equiv (u_k + v_k)\sqrt{N_0} V_\bk \times \exp{\bigl( ikR_0 \cos(\omega_m t -\phi) \bigr)}$. In Eq. \ref{Aeq_Ak(t)_prime} we make use of the Jacobi Anger expansion:
\begin{equation} \label{Aeq_jacobi_anger}
e^{i x \cos(\theta)} = \sum_{\nu=-\infty}^\infty i^\nu J_\nu(x) e^{i\nu\theta}.
\end{equation}
$J_\nu(x)$ is the Bessel function of the first kind of order $\nu$. After using Eq. \ref{Aeq_jacobi_anger} in Eq. \ref{Aeq_Ak(t)_prime}, we get
\begin{align} \label{eq_Akprime}
A^\prime_{\bf k}(t) &= -\frac{i}{\hbar}(u_k + v_k)\sqrt{N_0} V_\bk \sum_\nu i^\nu e^{-i\nu\phi} J_\nu(kR_0) \nonumber \\
& \quad \times \int \mathrm{d}t \, e^{-i(\omega_k -\nu \omega_m)t}. 
\end{align}
With the initial condition $V(t<0)=0$ and for a finite stirring time $t$, this gives
\begin{align} \label{Aeq_Akprime_sol}
 A^\prime_{\bf k}(t) &= -\frac{2i}{\hbar} (u_k + v_k) \sqrt{N_0} V_\bk \sum_{\nu} i^\nu e^{-i\nu\phi}  J_\nu(kR_0) \nonumber \\ 
 & \quad \times \frac{\sin\bigl[(\omega_k -\nu \omega_m)t/2 \bigr] }{ (\omega_k -\nu\omega_m)} e^{-i(\omega_k -\nu\omega_m)t/2}.
\end{align} 

 The energy of the system due to the perturbation at time $t$ is (see Eq. \ref{Aeq_energyExpectationT})
\begin{equation} \label{Aeq_SecondEnergy_circ}
\langle\Delta E(t)\rangle = \sum_{\bf k} \hbar\omega_k  |A^\prime_{\bf k}(t)|^2.
\end{equation}
Replacing $|A^\prime_{\bf k}(t)|^2$ using Eq. \ref{Aeq_Akprime_sol}, which gives
\begin{align}\label{Aeq_SecondEnergy_circ2}
& \langle\Delta E(t)\rangle = \frac{4}{\hbar} \sum_{\bf k}\omega_k (u_k + v_k)^2 N_0 |V_\bk|^2 \sum_{\nu, \nu^\prime} 
   \nonumber \\ 
& \quad  \times i^{\nu} (-i)^{\nu^\prime} e^{i(\nu-\nu^\prime)(\omega_m t/2-\phi)}  J_\nu(kR_0) J_{\nu^\prime}(kR_0) \nonumber \\ 
& \quad  \times \frac{\sin\bigl[(\omega_k -\nu\omega_m)t/2 \bigr] }{ (\omega_k -\nu\omega_m)}  \frac{\sin\bigl[(\omega_k -\nu^\prime\omega_m)t/2 \bigr] }{ (\omega_k -\nu^\prime\omega_m)}.
\end{align}
We use $\sum_{k_z}\delta_{k_z}=1$ in Eq. \ref{Aeq_SecondEnergy_circ2} and that reduces ${\bf k}=(k_x,k_y)$. In polar coordinates, we then integrate out the polar angle $\phi$ using $\int_0^{2\pi} \mathrm{d} \phi \exp[-i(\nu-\nu^\prime)\phi] = 2\pi \delta_{\nu^\prime,\nu}$. With this, we arrive at
\begin{align} \label{Aeq_circ_energySecond}
\langle\Delta E(t)\rangle &= \frac{2A }{\pi \hbar} \int_0^\infty \mathrm{d}k \, k \omega_k  (u_k + v_k)^2 N_0 |V_\bk|^2  \sum_{\nu} J_\nu^2(kR_0) \nonumber \\ 
& \quad  \times   \frac{\sin^2\bigl[ (\omega_k -\nu\omega_m)t/2 \bigr] }{(\omega_{k} -\nu\omega_m)^2}.
\end{align}
In the $t\rightarrow \infty$ limit the steady value of the energy in Eq. \ref{Aeq_circ_energySecond} is proportional to $t$. Therefore, the heating rate is  
\begin{align} \label{Aeq_circ_heatingRateDelta}
\frac{dE}{dt}   &= \frac{A }{\hbar} \sum_{\nu} \int_0^\infty \mathrm{d}k \, k \omega_k (u_k + v_k)^2 N_0 |V_\bk|^2 J_\nu^2(kR_0) \nonumber \\ 
& \quad  \times \delta(\omega_{k} -\nu\omega_m).
\end{align}
We solve the $\delta$-function in Eq. \ref{Aeq_circ_heatingRateDelta} for variable $k$ using the property $\delta(f(x))=\delta(x-x_0)/|f^\prime(x_0)|$. With this, we obtain
\begin{align} \label{Aeq_circ_heatingRate1}
\frac{dE}{dt}   &= \frac{A}{2\hbar}  \sum_{\nu} k_0 \omega_{k_0}  (u_{k_0} + v_{k_0})^2 N_0 |V_{k_0}|^2  J_\nu^2(k_0R_0) \nonumber \\ 
& \quad  \times \frac {\sqrt{(2m v_{\mathrm{B}})^2+(\hbar k_0)^2}} {2(mv_{\mathrm{B}})^2+(\hbar k_0)^2},
\end{align}
with $k_0=\bigl( \sqrt{4(m v_{\mathrm{B}})^4+(2m\hbar \nu\omega_m)^2} -2(m v_{\mathrm{B}})^2 \bigr)^{1/2}/\hbar$.
We now approximate the sum over the Bessel function index $\nu$ by an integral over a continuous variable in Eq. \ref{Aeq_circ_heatingRateDelta}, and then solve the $\delta$-function. This gives
\begin{align} \label{Aeq_circ_heatingRate2}
\frac{dE}{dt} & = \frac{A }{\hbar \omega_m} \int_0^\infty \mathrm{d}k \, k \omega_k  (u_k + v_k)^2 N_0 |V_\bk|^2 \nonumber \\ 
&\quad \times J_{\omega_k/\omega_m}^2(kR_0).
\end{align}

  We now express the heating rate for a lattice system. We therefore rewrite the potential in Eq. \ref{Aeq_potential_circFT} explicitly with $k_x, k_y$ as    
\begin{align} \label{Aeq_FourierVstir_Lattice}
&V_\bk(t) = V_\bk \nonumber \\
& \times \exp\bigl[ i \sqrt{k_x^2+k_y^2} R_0 \cos\bigl( \omega_m t-\tan^{-1}(k_y/k_x) \bigr) \bigr].
\end{align}
With this, the energy change in Eq. \ref{Aeq_SecondEnergy_circ} at time $t$ is
\begin{align} \label{Beq_energyLattice}
& \frac{\langle\Delta E(t)\rangle}{N_0V_0^2} =  \frac{8\pi^2 \sigma^4}{mA^2} \sum_{\bf k} \delta_{k_z} k^2 \exp(-k^2 \sigma^2) \sum_{\nu, \nu^\prime}   J_\nu(kR_0) \nonumber \\ 
&   \times  J_{\nu^\prime}(kR_0) \cos \Bigl[ (\nu-\nu^\prime) \Bigl( \frac{\pi}{2}+ \frac{\omega_m t}{2} -\tan^{-1}(k_y/k_x) \Bigl) \Bigr]   \nonumber \\ 
&  \times \frac{\sin\bigl[(\omega_k -\nu\omega_m)t/2 \bigr] }{ (\omega_k -\nu\omega_m)}  \frac{\sin\bigl[(\omega_k -\nu^\prime\omega_m)t/2 \bigr] }{ (\omega_k -\nu^\prime\omega_m)}.
\end{align}
In Eq. \ref{Beq_energyLattice} we used the explicit forms of $(u_k + v_k)^2$ and $|V_\bk|^2$. The time derivative of Eq. \ref{Beq_energyLattice} gives the heating rate, which can be written as
\begin{align}\label{Beq_eRateLattice}
R_{\mathrm{sc}} &= \frac{4\pi^2 \sigma^4}{mA^2} \sum_{\bf k} \delta_{k_z} k^2 \exp(-k^2 \sigma^2) \sum_{\nu}  J_\nu^2(kR_0) \nonumber \\
& \quad \times \frac{\sin\bigl[(\omega_k -\nu\omega_m)t \bigr] }{ (\omega_k -\nu\omega_m)} + R_{\mathrm{sc}}(\nu \neq \nu^\prime).
\end{align}
At long times, the heating rate is dominated by the diagonal contributions, i.e. the terms with $\nu = \nu'$. We evaluate this part of the sum in Eq. \ref{Beq_eRateLattice} for the first Brillouin zone, and show this analytical prediction by the dotted line in Fig. \ref{fig_circ_lin_comp}(b). 



\appendix


\begin{thebibliography}{999}
\bibitem{remark} More details on its derivation are given in Appendix \ref{App_eRate_lin}.
\bibitem{Stirrer_density} We estimate the density at the stirrer location using the Thomas-Fermi approximation of trapped system at $R_0=10\, \mu \mathrm{m}$.
\end{thebibliography}
\end{document}